\documentclass[12pt]{article}
\usepackage{natbib}
\bibpunct{(}{)}{,}{a}{}{;}
\usepackage[a4paper]{geometry}        
\usepackage{setspace}
\usepackage{amsmath,amsthm}
\usepackage{color,subfigure,graphicx}
\usepackage{url}
\usepackage{bm}
 \bmdefine\g{g} \bmdefine\K{K}\bmdefine\X{X}
\bmdefine\X{X} \bmdefine\D{D} \bmdefine\K{K} \bmdefine\Z{Z}
\bmdefine\x{x} \bmdefine\z{z} \bmdefine\y{y} \bmdefine\Y{Y}
\bmdefine\bfalpha{\alpha} \bmdefine\bfmu{\mu} \bmdefine\M{M}
\bmdefine\Q{Q} \bmdefine\P{P} \bmdefine\w{w} \bmdefine\W{W}
\bmdefine\p{p} \bmdefine\T{T} \bmdefine\t{t} \bmdefine\r{r}
\bmdefine\B{B} \bmdefine\I{I} \bmdefine\u{u} \bmdefine\p{p}
\bmdefine\Sig{\Sigma} \bmdefine\E{E} \bmdefine\F{F} \bmdefine\S{S}
\bmdefine\s{s} \bmdefine\w{w} \bmdefine\b{b} \bmdefine\W{W}
\bmdefine\w{w} \bmdefine\V{V} \bmdefine\v{v} \bmdefine\q{q}
\bmdefine\R{R} \bmdefine\A{A} \bmdefine\bflambda{\lambda}
\bmdefine\C{C} \bmdefine\U{U} \bmdefine\L{L} \bmdefine\d{d}
\bmdefine\c{c} \bmdefine\e{e} \bmdefine \H{H} \bmdefine\k{k}\bmdefine\N{N}
\newcommand{\bbeta}{\mbox{\boldmath$\beta$}}

\newcommand{\PP}{\mathcal{P}}

\newcommand{\dof}{\operatorname{DoF}}
\newcommand{\trace}{\operatorname{trace}}

\usepackage{amsmath,amssymb}
\newtheorem{conjecture}{Conjecture}

\newtheorem{propo}[conjecture]{Proposition}
\newtheorem{theo}[conjecture]{Theorem}
\newtheorem{defi}[conjecture]{Definition}
\usepackage{algorithmic}
\usepackage{algorithm}

\title{The Degrees of Freedom of \\Partial Least Squares Regression\thanks{to appear in the Journal of the American Statistical Association (2011)}}
\author{Nicole Kr\"amer\\Weierstrass Institute Berlin\\\texttt{nicole.kraemer@wias-berlin.de} \and Masashi Sugiyama\\Tokyo Institute of Technology\\\texttt{sugi@cs.titech.ac.jp}}
\begin{document}

\maketitle

\begin{abstract}
The derivation of statistical properties for Partial Least Squares regression can be a challenging task. The  reason is that the  construction of latent components from the predictor variables  also depends on the response variable. While this typically leads to good performance and interpretable models in practice, it makes the statistical analysis more involved. In this work, we study the intrinsic complexity of  Partial Least Squares Regression. Our contribution is an unbiased estimate of its Degrees of Freedom. It is defined as the trace of the first derivative of the fitted values, seen as a function of the response.  We establish two equivalent representations that rely on the close connection of Partial Least Squares to matrix decompositions and Krylov subspace techniques. We show that the Degrees of Freedom depend on the collinearity of the predictor variables: The lower the collinearity is, the higher the Degrees of Freedom are. In particular, they are typically higher than the naive approach that defines the Degrees of Freedom as the number of components. Further, we illustrate how our Degrees of Freedom estimate can be used for the comparison of different regression methods. In the experimental section, we show that our Degrees of Freedom estimate in combination with information criteria is useful for model selection.
\end{abstract}
\section{Introduction}
Partial Least Squares regression (PLSR) \citep{Wold7501} is a two-step regularized regression technique. It iteratively
constructs an orthogonal set of  latent components from the predictor variables  which have maximal
covariance with the response variable. This low-dimensional representation of the data  is then used to fit a linear regression model. PLSR is extended to nonlinear regression problems via a transformation of the predictor variables \citep{Durand97,Rosipal0101}.

For model selection in PLSR, the optimal number  of components has to be determined. While cross-validation is the standard approach, an alternative is the use of information criteria, which use the complexity of the fitted model.   In regression, the
complexity of a fitting method is defined in terms of its Degrees of Freedom. Apart from their usefulness for model selection, Degrees of Freedom  also quantify the intrinsic complexity of a regression method (see e.g. \cite{Voet99} for an overview).  In contrast to other standard regression techniques as Principal Components Regression or Ridge Regression (where the Degrees of Freedom equal the trace of the hat-matrix), the derivation of the Degrees of Freedom for PLSR is not straightforward. This is due to the fact that PLSR is not
linear in the sense that the fitted response does not depend linearly on the response. As the set of latent components is constructed in a supervised fashion, the projection of the response variable onto these components is a highly nonlinear function of the response. Therefore, it has been argued (e.g \cite{Martens8901,Frank9301}) that the Degrees of Freedom of PLSR exceed the number  of components.

We provide an unbiased estimate of the generalized Degrees of Freedom of PLSR. It is defined as the trace of the Jacobian matrix of the fitted values, seen as a function of the response.  We illustrate on benchmark data that the complexity of PLSR depends on  the collinearity of the predictor variables: The higher the collinearity is, the lower the complexity is. Under additional assumptions on the collinearity structure of the data, we provide bounds for the Degrees of Freedom if one component is used.

We present two different implementations. (i) The first one is derived via an  iterative computation of the first derivative of the PLSR fit. To do so, we use the equivalence of Partial Least Squares Regression to the Lanczos decomposition of the matrix of predictor variables. This implementation has the advantage that it also provides an asymptotic distribution of the PLSR regression coefficients, which can be used for the construction of confidence intervals \citep{Phatak0201,Denham9701}.  (ii) The second implementation computes the Degrees of Freedom directly, i.e. it avoids the computation of the derivative itself. This leads to a more favorable runtime. For the derivation, we use the close connection of  PLS regression to Krylov subspace techniques. Both algorithms are implemented in the R package `plsdof' \citep{plsdof}.

We investigate the performance of the Degrees of Freedom of PLSR with respect to model selection of the number of PLSR components. We  compare the test errors based on 10-fold cross-validation and based on the Bayesian Information Criterion (BIC) \citep{Schwa78} in a simulation study. For the latter information criterion, we use our Degrees of Freedom estimate and the naive approach that defines the Degrees of Freedom of PLSR via the number of components. Our experiments show that the model selected based on our Degrees of Freedom estimate is typically less complex than the model selected by the naive approach. This can lead to a higher test error for the naive approach. In terms of prediction accuracy, our Degrees of Freedom approach is  on a par with the gold-standard of cross-validation, providing further evidence that our estimates captures the true model complexity correctly.

\section{Methodological Background} \label{sec:pls}
We consider a multivariate regression problem
\begin{eqnarray}
\label{eq:reg1}
\mathbb{R} \ni Y_i&=& f(\x_i) + \varepsilon_i\,,\,\varepsilon_i \sim \mathcal{N}\left(0,\sigma^2\right)\,,
\end{eqnarray}
and the task is to estimate the unknown function $f:\mathbb{R}^p\rightarrow\mathbb{R}$ from a finite set of $n$ examples $(\x_1,y_1),\ldots,(\x_n,y_n) \in \mathbb{R}^p \times \mathbb{R}$, where $y_i$ is  drawn from \eqref{eq:reg1}. We assume that the error variables $\varepsilon_i$ are independent. Let us denote by $\overline{\x}$ and $s(\x)$ the mean and the standard deviation of the predictor examples $\x_i$ and by $\overline{\y}$ the mean of the response samples $y_i$.  The $n \times p$ data matrix $\X$ is the matrix whose rows are the centered and scaled (to unit variance) $\x_i$, and the vector $\y \in
\mathbb{R}^n$ consists of the centered response $y_i$. While in the course of this paper, we assume that the regression function $f$ is linear,
\begin{eqnarray}
\label{eq:linmodel}
f(\x)&=& \beta^{(0)} + \bbeta^\top \x\,,\,\beta^{(0)} \in \mathbb{R}, \bbeta \in \mathbb{R}^p\,,
\end{eqnarray}
the definitions given in Subsection \ref{subsec:DoF} do not require $f$ to be a linear function. Finally, we define
\begin{eqnarray}
\label{eq:scatter} \S= \frac{1}{n-1} \X^\top \X \in \mathbb{R}^{p \times p}&\text{ and }&\s= \frac{1}{n-1}\X^\top \y \in \mathbb{R}^p.
\end{eqnarray}
\subsection{Degrees of Freedom and Model Selection} \label{subsec:DoF}                                  %
Regularized regression methods typically yield a set of  estimates $\widehat f_{\lambda}$ of the true regression function $f$, and the parameter $\lambda$ determines the amount of regularization. In Partial Least Squares Regression, the parameter $\lambda$ corresponds to the number of latent components.

The task is to determine the optimal parameter value $\lambda$. Information criteria are based on the rationale that the true error of $\widehat f_{\lambda}$ can be estimated in terms of its training error and its complexity. In regression problems, the complexity is defined via  Degrees of Freedom. These are defined for the class of methods that are linear in the sense that the fitted values are a linear function of $\y$, i.e. $\widehat \y_{\lambda}=\H_{\lambda} \y$ with $\H_{\lambda} \in \mathbb{R}^{p \times p}$ a matrix that does not depend on $\y$. Popular examples are Ridge Regression and Principal Components Regression. The matrix $\H_{\lambda}$ is called the hat-matrix. In the linear case, the Degrees of Freedom are defined as the trace of the hat-matrix,
 \begin{eqnarray}
 \label{eq:hat}
\dof(\lambda)&=& \text{trace} \left(\H_{\lambda}\right)\,.
\end{eqnarray}
As we point out below, PLS regression is not a linear method, and the above definition cannot be applied. In order to extend the notion of Degrees of Freedom to PLSR, we employ the generalized definition proposed by
 \cite{Efron0401}.
 \begin{defi}\label{def:dof}
 Let $\widehat f_{\lambda}$ be an estimate of the true regression function $f$, parameterized by $\lambda$. We define the vector of fitted values as $\widehat \y_{\lambda}=(\widehat f_{\lambda}(\x_1,),\ldots,\widehat f_{\lambda}(\x_n))^\top$. The Degrees of Freedom are
 \begin{eqnarray*}
 \dof\left(\lambda\right)&=& \mathbb{E}\left[\trace\left( \frac{\partial \widehat \y_{\lambda}}{\partial \y}\right)   \right]\,.
\end{eqnarray*}
Here, the input $\X$ is assumed to be fixed and the expectation $\mathbb{E}$ is taken with respect to $y_1,\ldots,y_n$.
\end{defi}
The Degrees of Freedom measure the sensitivity of the fitted values, seen as a function of $\y$. Note that for the special case of linear methods, this definition coincides with \eqref{eq:hat}.

Popular examples of information criteria include  the Akaike information criterion  \citep{Aka73}, the generalized minimum description length \citep{HanYu2001} and
 the Bayesian information criterion (BIC) \citep{Schwa78}
 \begin{eqnarray*}
 \label{eq:bic}
\text{bic}\left(\lambda\right)&=& \left\|\widehat \y_{\lambda} -\y\right \|^2 + \log(n)\sigma^2 \dof(\lambda)\,.
\end{eqnarray*}
For information criteria, we need an estimate of the noise level $\sigma$ defined in \eqref{eq:reg1}. For linear methods $\widehat \y_{\lambda}= \H_{\lambda} \y$, this is accomplished as follows. The residual is of the form $\y - \widehat \y_{\lambda}= \left(\I_n - \H_{\lambda} \right)\y.$ The bias-variance decomposition of the mean squared error yields
\begin{eqnarray*}
E\left[\|\y - \widehat \y_{\lambda} \|^2\right]
&=& \left \|E[ \y - \widehat \y_{\lambda} ]\right\|^2 +  \operatorname{trace}\left( (\I_n-\H_{\lambda})(\I_n-\H_{\lambda} ^\top)\right) \sigma^2\,.
\end{eqnarray*}
 By dropping the unknown bias term $\left \|E[ \y - \widehat \y_{\lambda} ]\right\|^2$, we yield an estimate of $\sigma$ via
\begin{eqnarray}
\label{eq:sigmastar}
\widehat \sigma_* ^2&=& \frac{\|\widehat \y_{\lambda} - \y \|^2}{\operatorname{trace}\left( (\I_n-\H_{\lambda})(\I_n-\H_{\lambda} ^\top)\right)}\,.
\end{eqnarray}
If $\H_{\lambda}$ is a projection operator (which is e.g. true for Principal Components Regression), the expression is simplified  to
\begin{eqnarray}
\label{eq:sigma}
\widehat \sigma ^2&=& \frac{\|\widehat \y_{\lambda} - \y \|^2}{n- \dof(\lambda)}\,.
\end{eqnarray}
We note that the latter estimate is  most commonly used, even if the above assumption is not fulfilled.
\subsection{Partial Least Squares Regression}\label{subsec:pls} %
PLSR constructs $m$  latent components  $\T=\left(\t_1,\ldots,\t_m\right) \in \mathbb{R}^{n \times m}$ from the predictor variables $\X$ such that
the components $\t_i$ are mutually orthogonal and that they have maximum covariance to the response $\y$. In the NIPALS algorithm \citep{Wold7501}, the first component $\t_1= \X \w_1$ maximizes the squared covariance to the response $\y$,
\begin{eqnarray}
\label{eq:critneu} \w_1&=&\text{arg}\max_{\w}
\frac{\left\|\text{cov}(\X\w,\y)\right\|^2}{\w^\top
\w}=\text{arg}\max_{\w} \frac{\w^\top \X^\top \y \y^\top \X \w}{\w^\top \w}\propto \X^\top \y\,.
\end{eqnarray}
Subsequent components $\t_2,\t_3,\ldots$ are chosen such that they
maximize the squared covariance to $\y$  and that all components are
mutually orthogonal. Orthogonality is enforced by  deflating the
original variables $\X$\,,
\begin{eqnarray}
\label{eq:deflation} \X_i &=& \X - \PP_{\t_1,\ldots,\t_{i-1}} \X \,.
\end{eqnarray}
Here, $\PP_{\t_1,\ldots,\t_{i-1}}$ denotes the orthogonal projection onto the space spanned by $\t_1,\ldots,\t_{i-1}$. We then replace $\X$ by $\X_i$  in (\ref{eq:critneu}). While the matrix $\T=\left(\t_1,\ldots,\t_m\right)$ is orthogonal by construction, it can be shown that the matrix $\W=(\w_1,\ldots,\w_m)\in \mathbb{R}^{d\times m}$ is orthogonal  as well (e.g. \cite{Hoskuldsson8801}).  The $m$ latent components $\T$ are used as regressors in a least squares fit in place of $\X$, leading to fitted values
\begin{eqnarray}
\label{eq:fit}
\widehat \y_m&=&\overline{\y} {\bf 1}_{n}+  \T \left(\T ^\top \T\right)^{-1} \T ^\top \y=\overline{\y}{\bf 1}_{n} + \mathcal{P}_{\T} \y\,.
\end{eqnarray}
We emphasize again that PLSR is not a linear estimator as defined in Section \ref{subsec:DoF}. The projection matrix $\PP_{\T}$ depends on the response as well, as the latent components $\T$ are defined in terms of both $\X$ and $\y$.

To determine the estimated regression coefficients and intercept in \eqref{eq:linmodel}, we define
\begin{eqnarray}
\label{eq:L}
\L&=&  \T^\top \X \W \in \mathbb{R}^{m\times m}\,,
\end{eqnarray}
and obtain $\widehat \bbeta_m= \D \W \L^{-1} \T^\top \y$ for the regression coefficients \citep{Manne8701,Hoskuldsson8801} and $\widehat{\beta}^{(0)} _m= \overline{\y} - \overline{\x}^\top \widehat \bbeta_m$ for the intercept. Here, $\D$ is the diagonal $p \times p$ scaling matrix with entries $d_{ii}=1/s(\x)_i$.

\section{Unbiased Estimation of the Degrees of Freedom} %
The latent components $\T$ of PLSR depend on the response $\y$. Therefore, the relationship between $\y$ and the fitted PLSR values $\widehat \y_m$ is nonlinear, and the compact formula \eqref{eq:hat} for the Degrees of Freedom cannot be applied. However, we can use the more general Definition \ref{def:dof} to obtain an unbiased plug-in estimate.
\begin{propo}
An unbiased  estimate of the Degrees of Freedom of PLSR with $m$ latent components $\T=(\t_1,\ldots,\t_m)$ is given by
\begin{eqnarray}
\label{eq:dof}
  \widehat {\dof}(m) &=& 1+ \trace\left(
    \frac{\partial\PP_{\T} \y}{\partial \y}
  \right).
\end{eqnarray}
\end{propo}
The constant term $1$ corresponds to the estimation of the intercept $\beta^{(0)}$, which consumes one Degree of Freedom. For the derivation, we need to compute the trace of the derivative in \eqref{eq:dof} explicitly. We propose two equivalent algorithms in Subsections \ref{subsec:lanczos} and \ref{subsec:krylov}.
\subsection{Illustration and a Lower Bound}
\begin{table}
\caption{Properties of the three benchmark data sets. A detailed description of the data can be found in the appendix.
}
\begin{center}
\begin{tabular}{lrrl}
\hline
data set &variables&examples&mean absolute correlation\\
\hline
\texttt{kin (fh)}& $32$& $8192$&low ($0.009$) \\
\texttt{ozone} & $12$&$203$&medium ($0.260$)\\
\texttt{cookie}& $700$&$70$&high ($0.867$)\\
\hline
\end{tabular}
\end{center}

\label{tab:data}
\end{table}
Before delving deeper into the details of the implementation, we illustrate the properties of the Degrees of Freedom on benchmark data. An overview of the data sets is given in Table \ref{tab:data} (see the appendix for more details). We choose the three particular data sets as they differ with respect to the collinearity structure of the predictor variables. As an indicator for the degree of collinearity, we compute the mean of the absolute empirical correlation coefficients defined by $\overline{s}= (2/(p^2 -p))\sum_{i<j} ^{p} \left|s_{ij}\right|\,.$
Here $s_{ij}$ is the $(i,j)$-entry of the empirical correlation matrix  $\S$  of $\X$. The values of $\overline{s}$ are displayed in the fourth column of Table \ref{tab:data}.

\begin{figure}[htb]
 \centering{\includegraphics[width=0.22\textheight]{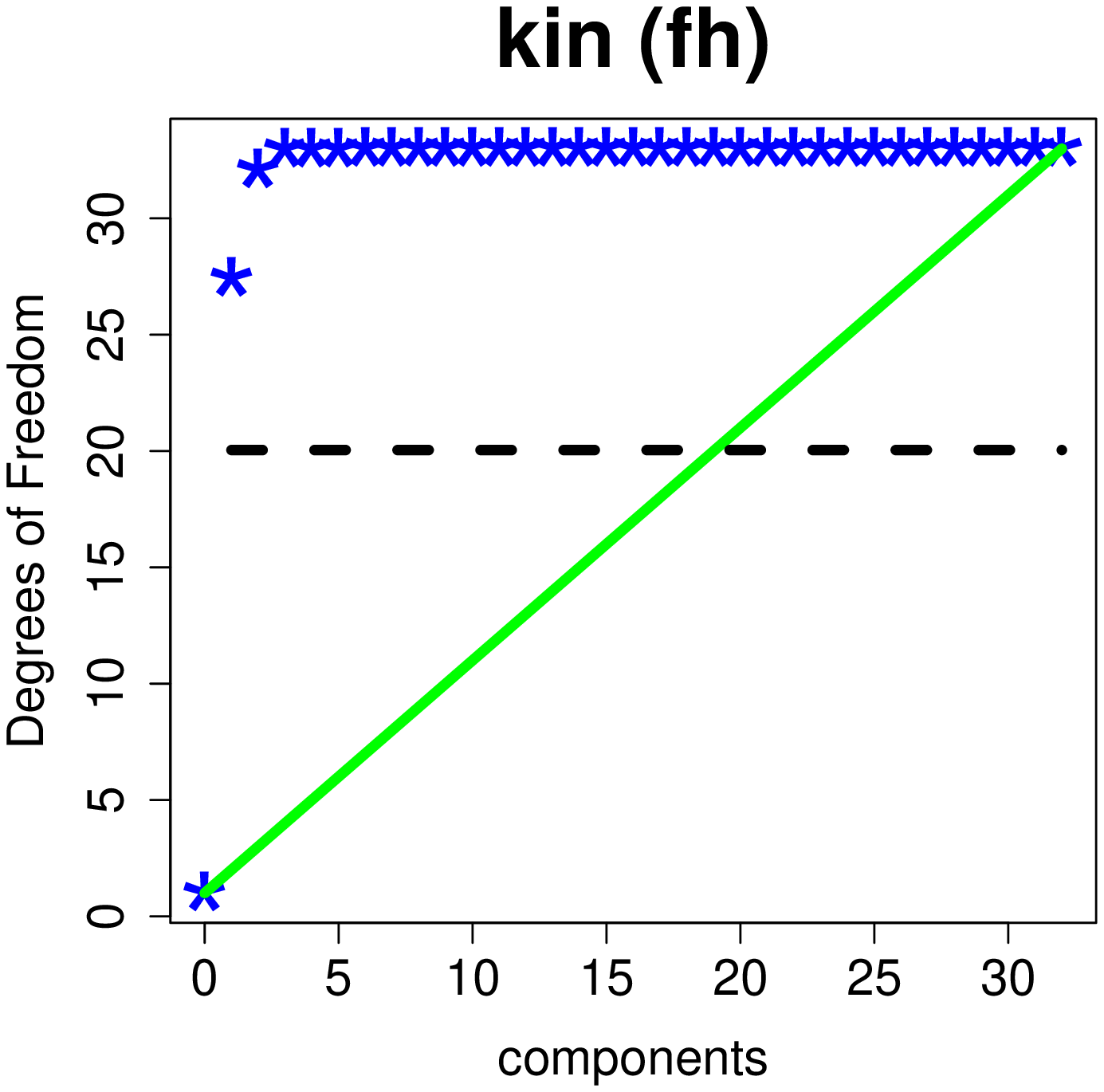} \includegraphics[width=0.22\textheight]{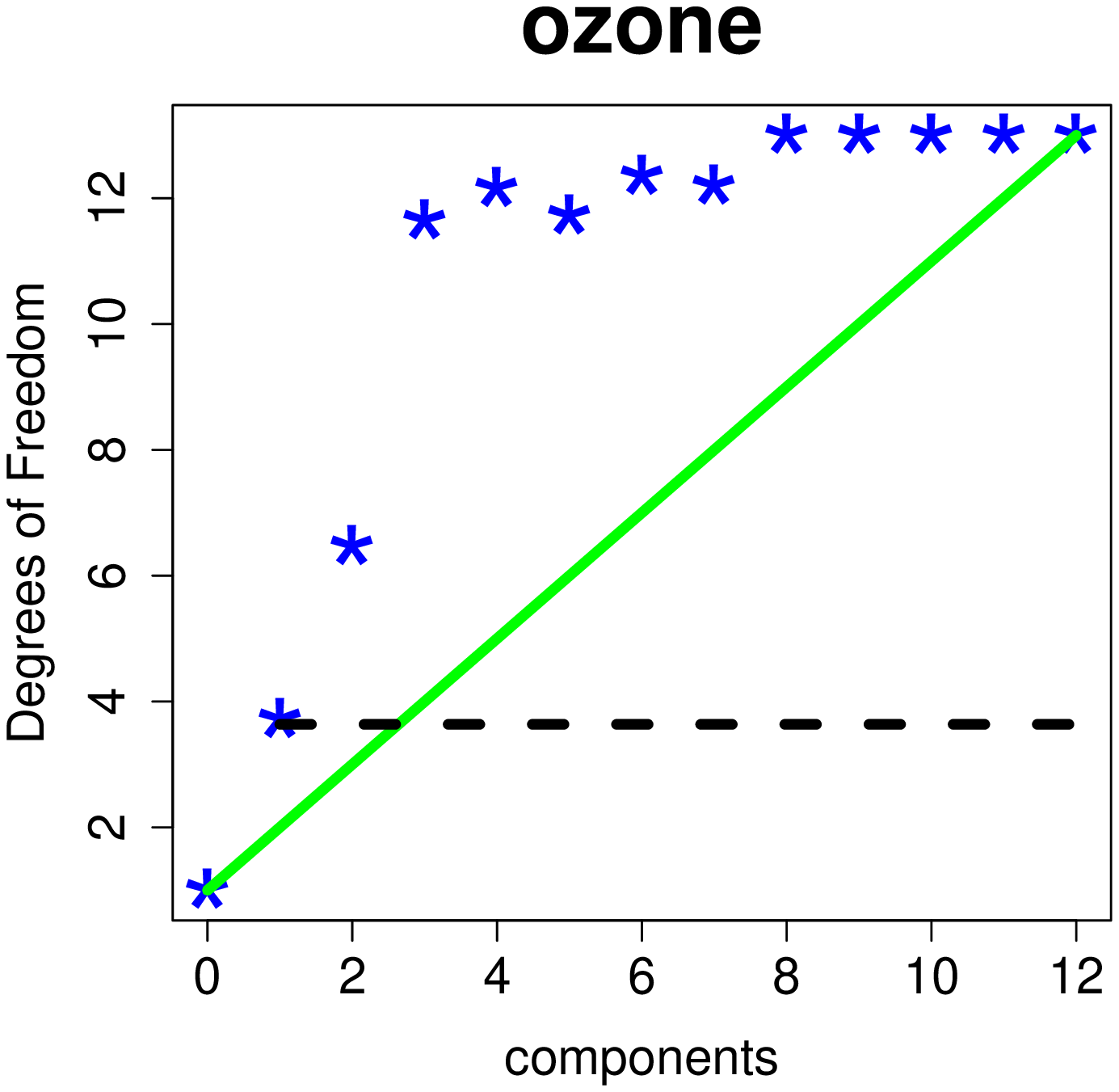} \includegraphics[width=0.22\textheight]{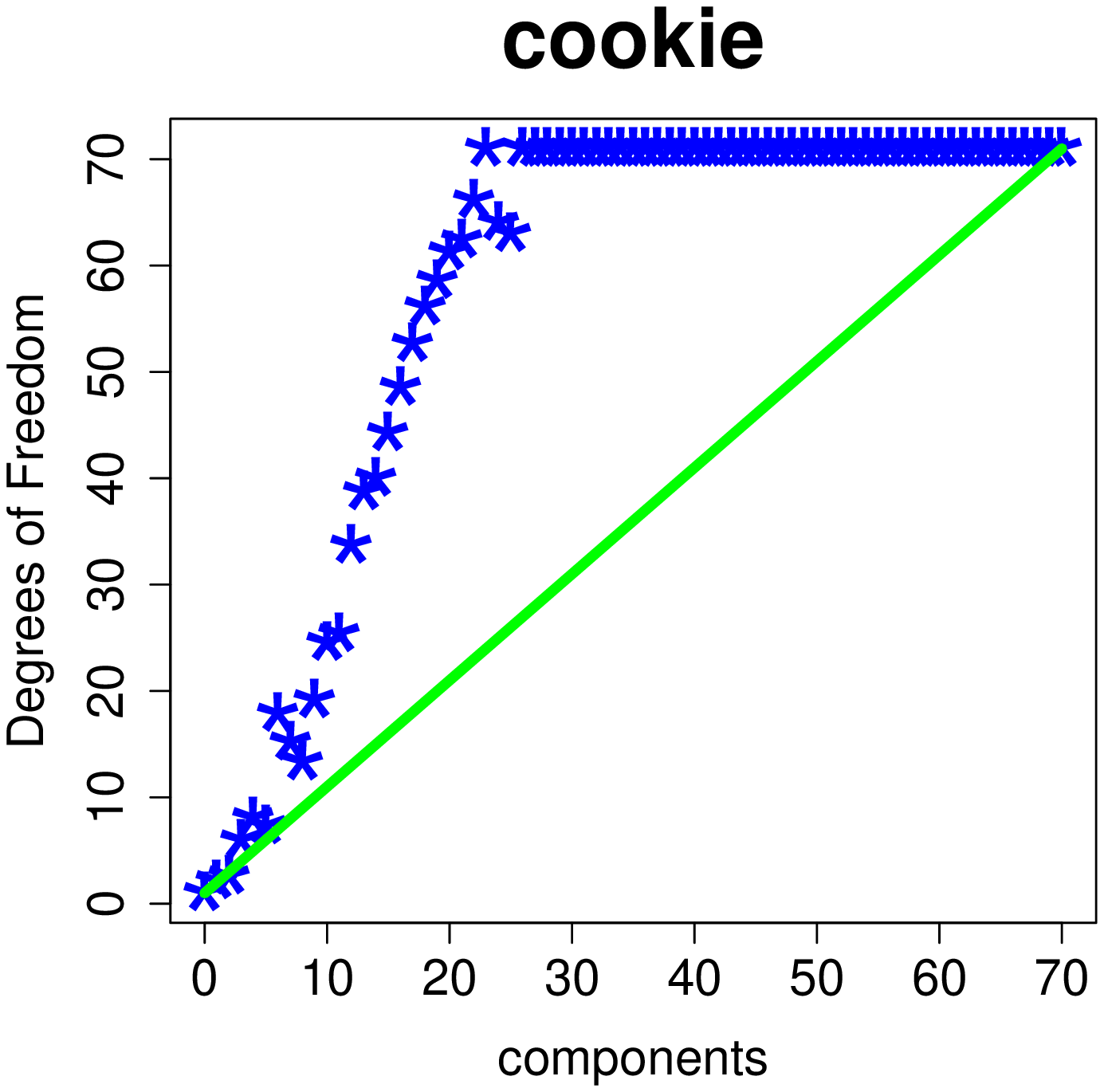}}
  \caption{Estimated Degrees of Freedom (stars) for the three benchmark data sets. The solid line displays the naive estimate $\dof(m)=m+1$. If the assumption of theorem \ref{thm:bound} is fulfilled, we also display the lower bound on the Degrees of Freedom for $1$ component (dashed horizontal line).}
  \label{fig:DoF}
\end{figure}
In Figure \ref{fig:DoF}, we plot the Degrees of Freedom for the three data sets as a function of the number of components. (For the large data set \texttt{kin (fh)}, we use a subsample of size $300$.) In addition, we display the naive estimate $\dof(m)=m+1$. In the examples, our Degrees of Freedom estimate always exceeds the naive approach. This supports the conjecture  that $\dof(m)\geq m+1$, which is voiced e.g. in \cite{Martens8901} and \cite{Frank9301}.

Furthermore, Figure \ref{fig:DoF} shows that the correlation structure of the predictor variables determines the shape of the DoF-curve. In our examples, the complexity is higher for data with low correlation. We underpin this observation with a lower bound on the Degrees of Freedom of PLSR  with $m=1$ component.
\begin{theo}\label{thm:bound}
If  the largest eigenvalue $\lambda_{\max}$  of the empirical correlation matrix $\S$ defined in \eqref{eq:scatter} fulfills
\begin{eqnarray}
\label{eq:lambdamax}
\lambda_{\max} &\leq & \frac{1}{2} \text{trace} (\S)\,,
\end{eqnarray}
then
\begin{eqnarray}
\label{eq:low}
\widehat{\dof} (m=1)&\geq &1+ \frac{\text{trace}(\S)}{\lambda_{\max}}\,.
\end{eqnarray}
\end{theo}
Condition \eqref{eq:lambdamax} controls the amount of  collinearity of $\X$. If the collinearity is low,  the decay of the eigenvalues of $\S$ is slow (and condition \eqref{eq:lambdamax} is fulfilled). The lower bound \eqref{eq:low} is higher for data with low collinearity. In Figure \ref{fig:DoF}, we add the lower bound for the data sets \texttt{kin (fh)} and \texttt{ozone}, which fulfill  condition \eqref{eq:lambdamax}.

\begin{proof}
We express the PLS fit for one component in terms of $\S$ and $\s$ defined in \eqref{eq:scatter}. Recall that the first latent component is defined as $\t_1= \X \s$, which implies
\begin{eqnarray*}
\widehat \y_1&=& \overline \y +  \frac{\s^\top \s}{\s^\top \S \s}\X \s\,.
\end{eqnarray*}
After computing the derivative of this term with respect to $\y$ and computing its trace, we obtain
\begin{eqnarray*}
\widehat \dof(m=1)&=& 3 + \frac{\s^\top  \s}{\s^\top \S \s} \left[\text{trace}(\S) - 2 \frac{ \left(\s^\top \S^2 \s \right)}{\s^\top \S \s}\right]\,.
\end{eqnarray*}
Now, by definition of maximal eigenvalues, $\s^\top \S^2 \s /\s^\top \S \s\leq \lambda_{max}$ and $\s^\top \s /\s^\top \S \s\geq 1/\lambda_{max}\,.$ It follows that
\begin{eqnarray*}
\text{trace}(\S) - 2 \frac{ \left(\s^\top \S^2 \s \right)}{\s^\top \S \s}&\geq& \text{trace}(\S) -2 \lambda_{\max}\,.
\end{eqnarray*}
Condition \eqref{eq:lambdamax} ensures that the right-hand side of this inequality is non-negative, hence,
\begin{eqnarray*}
\widehat \dof(m=1)&\geq&3 + \frac{\s^\top  \s}{\s^\top \S \s}\left[\text{trace}(\S) - 2 \lambda_{\max}\right]\\
&\geq& 3 + \frac{1}{\lambda_{max}}\left[\text{trace}(\S) - 2 \lambda_{\max}\right]= 1+ \frac{\text{trace}(\S)}{\lambda_{\max}}\,.
\end{eqnarray*}
\end{proof}
\subsection{Comparison of Regression Methods}\label{subsec:comparison}
While the regularization parameters $\lambda_1$ and $\lambda_2$  for two competing regression methods cannot be compared directly, their corresponding Degrees of Freedom values $\dof(\lambda_1)$ and $\dof(\lambda_2)$ are on the same scale. Hence, Degrees of Freedom allow us to compare the model complexity across different regularized regression approaches.

To illustrate this point, we compare the model complexity of PLSR, Principal Components Regression (PCR), and Ridge Regression on the \texttt{ozone} data set. Recall that for PCR, the Degrees of Freedom are the number of principal components plus one, and for Ridge Regression,
\begin{eqnarray*}
\dof(\lambda)_{\text{ridge}}&=& 1 + \text{trace} \left(\,\X \left(\X^\top \X + \lambda \I_p\right)^{-1} \X^\top\right) \,.
\end{eqnarray*}
We split the data set into $50$ training examples and $153$ test examples. On the training data, we compute the optimal model parameter for the respective methods with 10-fold cross-validation. We assess the predictive performance by computing the mean squared error of prediction on the test set. This procedure is repeated 50 times.

\begin{figure}[htb]
 \centering{\includegraphics[width=0.2\textheight]{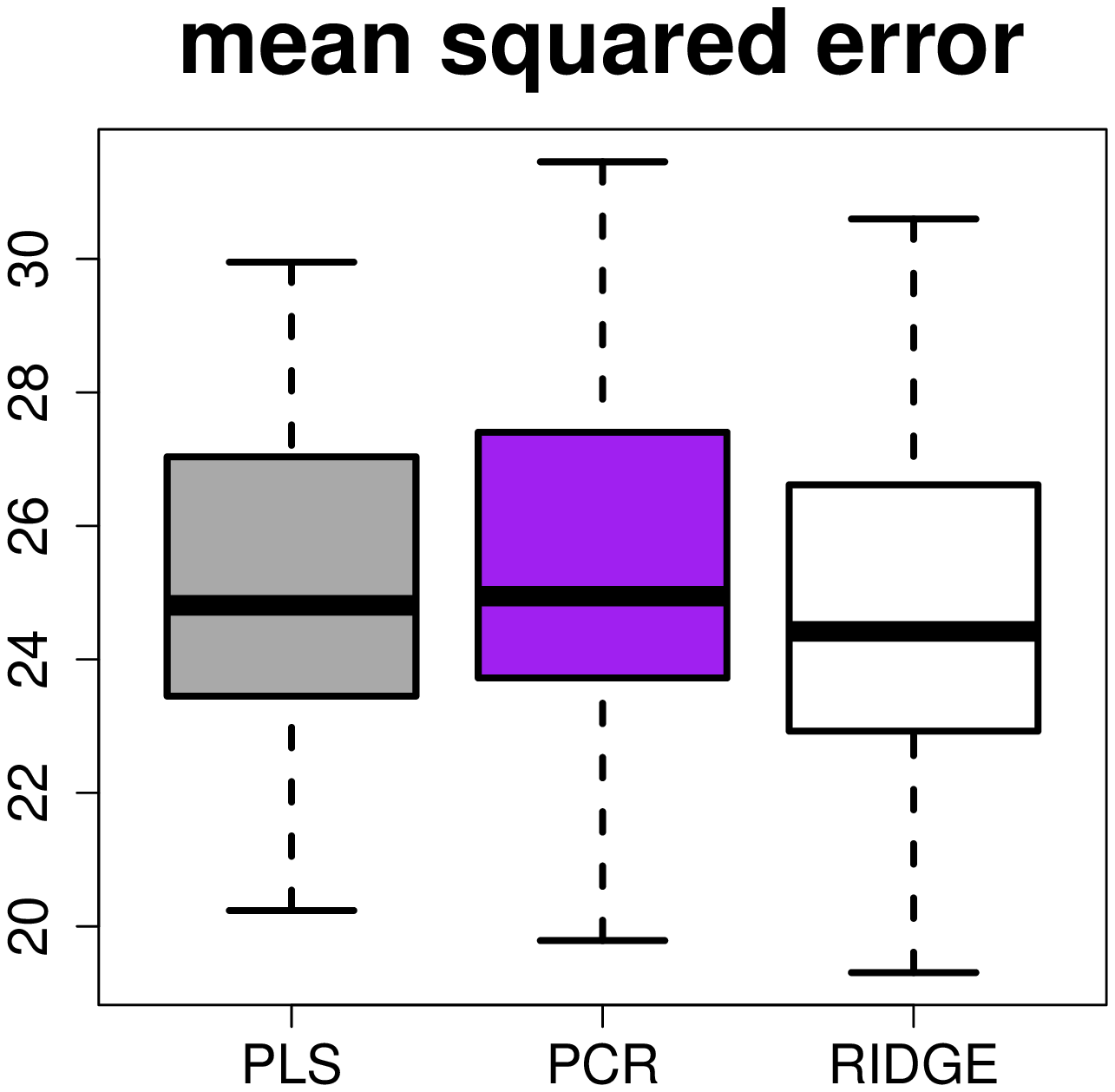} \includegraphics[width=0.2\textheight]{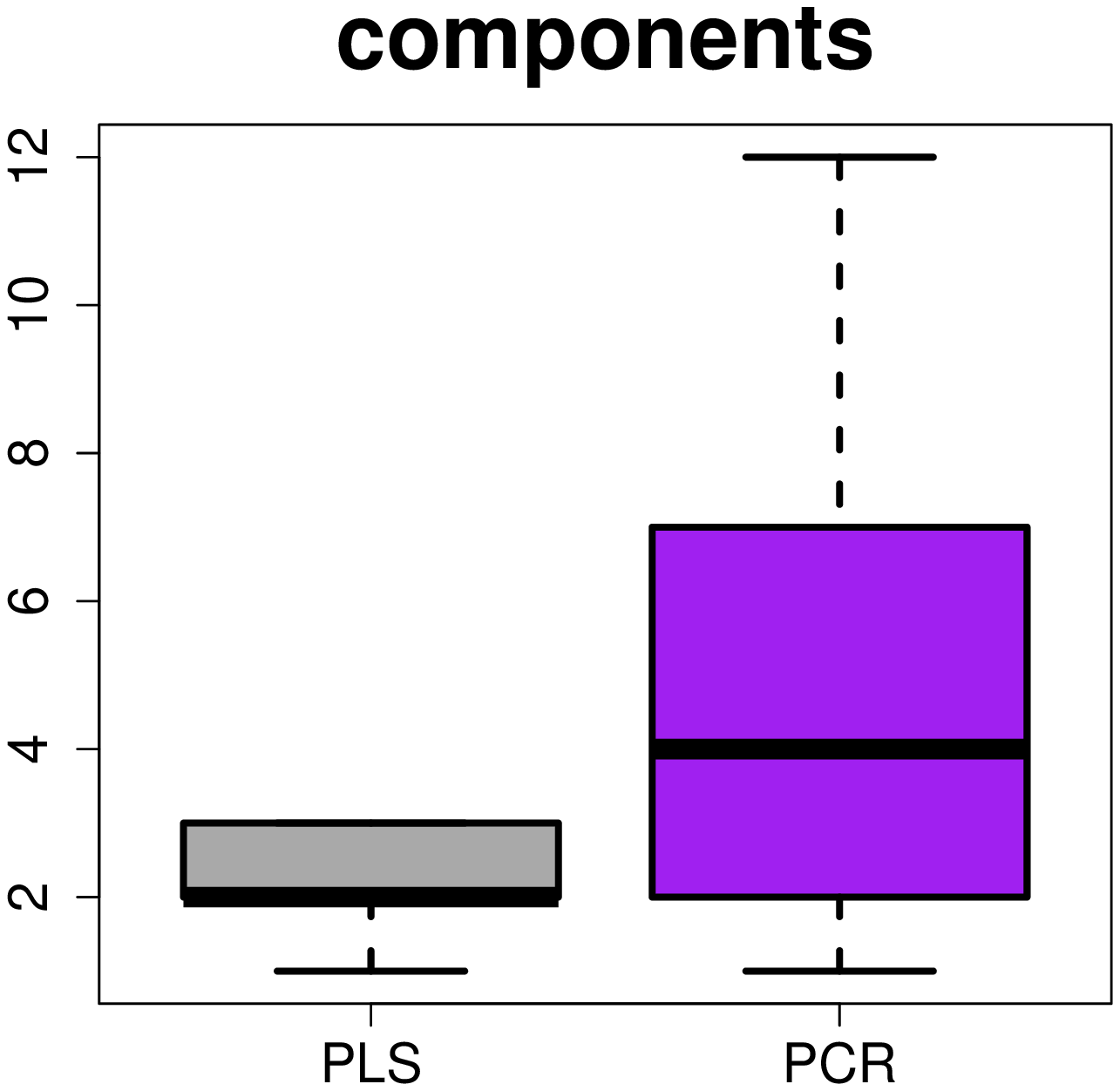} \includegraphics[width=0.2\textheight]{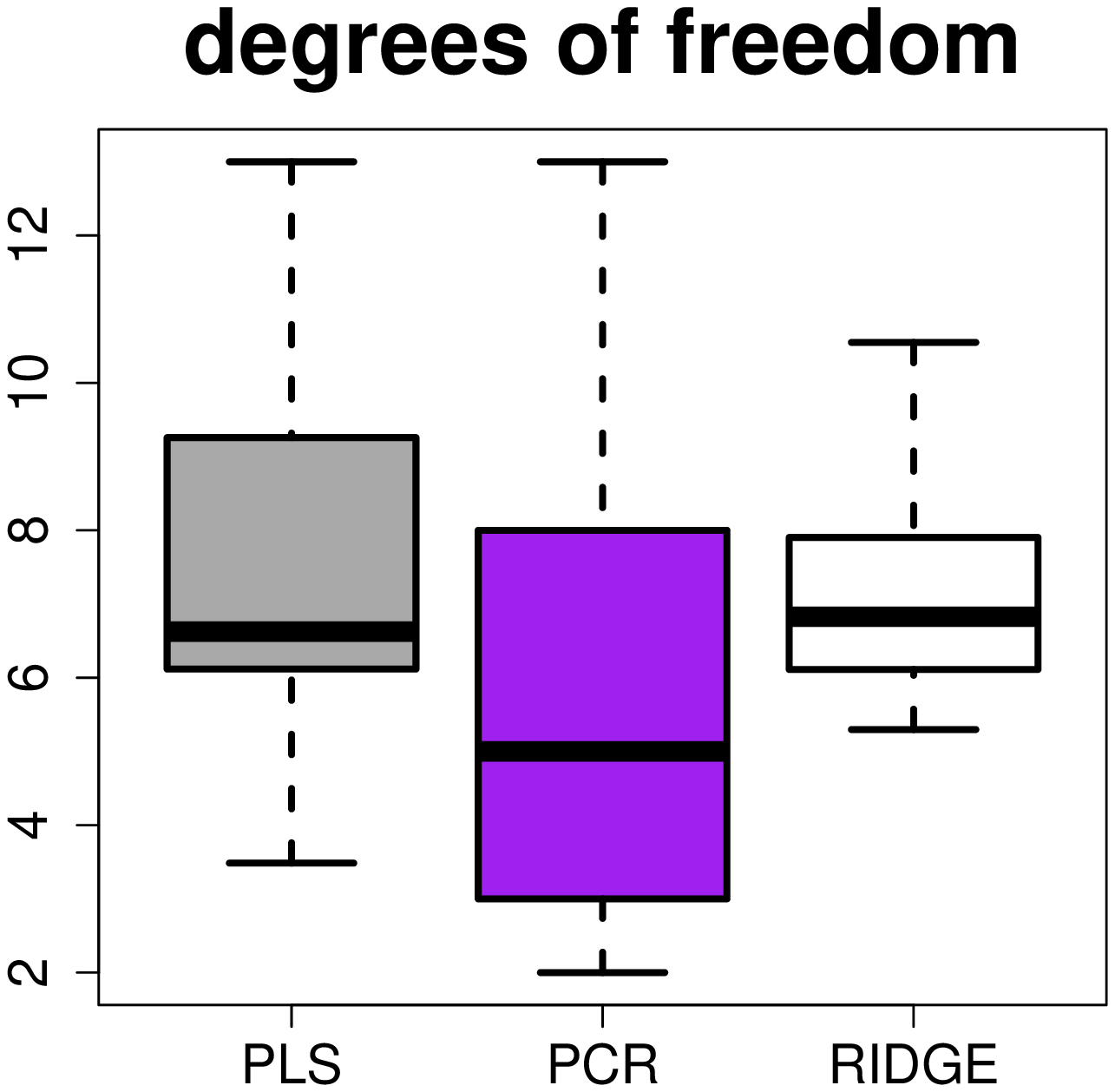}}
  \caption{Comparison of PLS, PCR and Ridge Regression. Left: mean squared error of prediction for the three regression methods. Center: Number of cross-validation optimal number of components for PLS and PCR. Right: Cross-validation optimal Degrees of Freedom for the three regression methods. }
  \label{fig:regression}
\end{figure}

Let us start with the observation that the predictive performance of the three methods is similar on this data set (left plot in Figure \ref{fig:regression}). This is in accordance with previous observations that the methods often  perform similarly (see e.g. \citet{Frank9301}). Next, we compare the model complexity of PLSR and PCR in terms of the cross-validation-optimal number of components. This is displayed in the center plot of Figure \ref{fig:regression}. On average, PLSR selects fewer components than PCR. However, in terms of Degrees of Freedom, PLSR selects more complex models than PCR (right plot in Figure \ref{fig:regression}).

\begin{figure}[htb]
 \centering{\includegraphics[width=0.3\textheight]{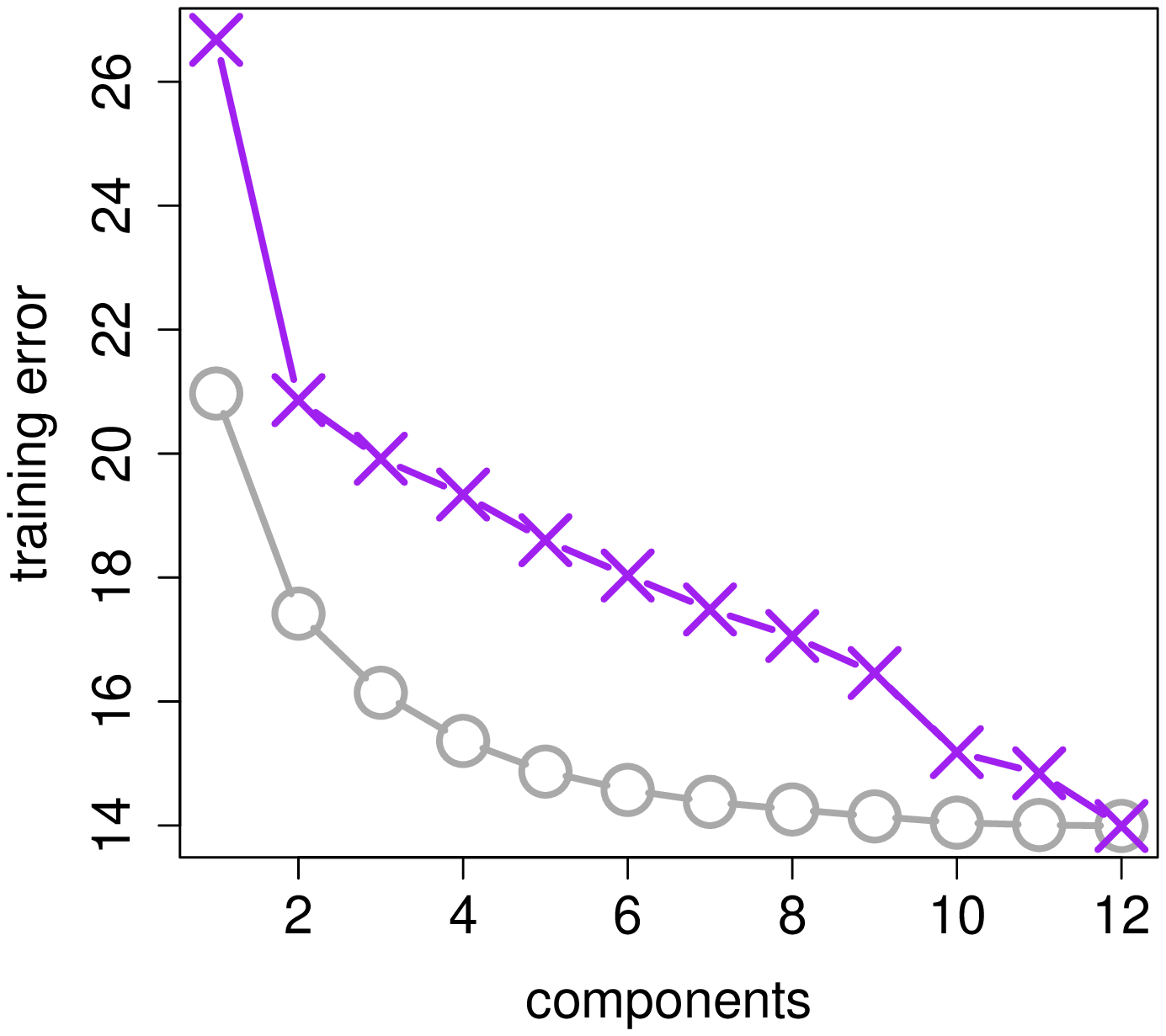}
 \includegraphics[width=0.3\textheight]{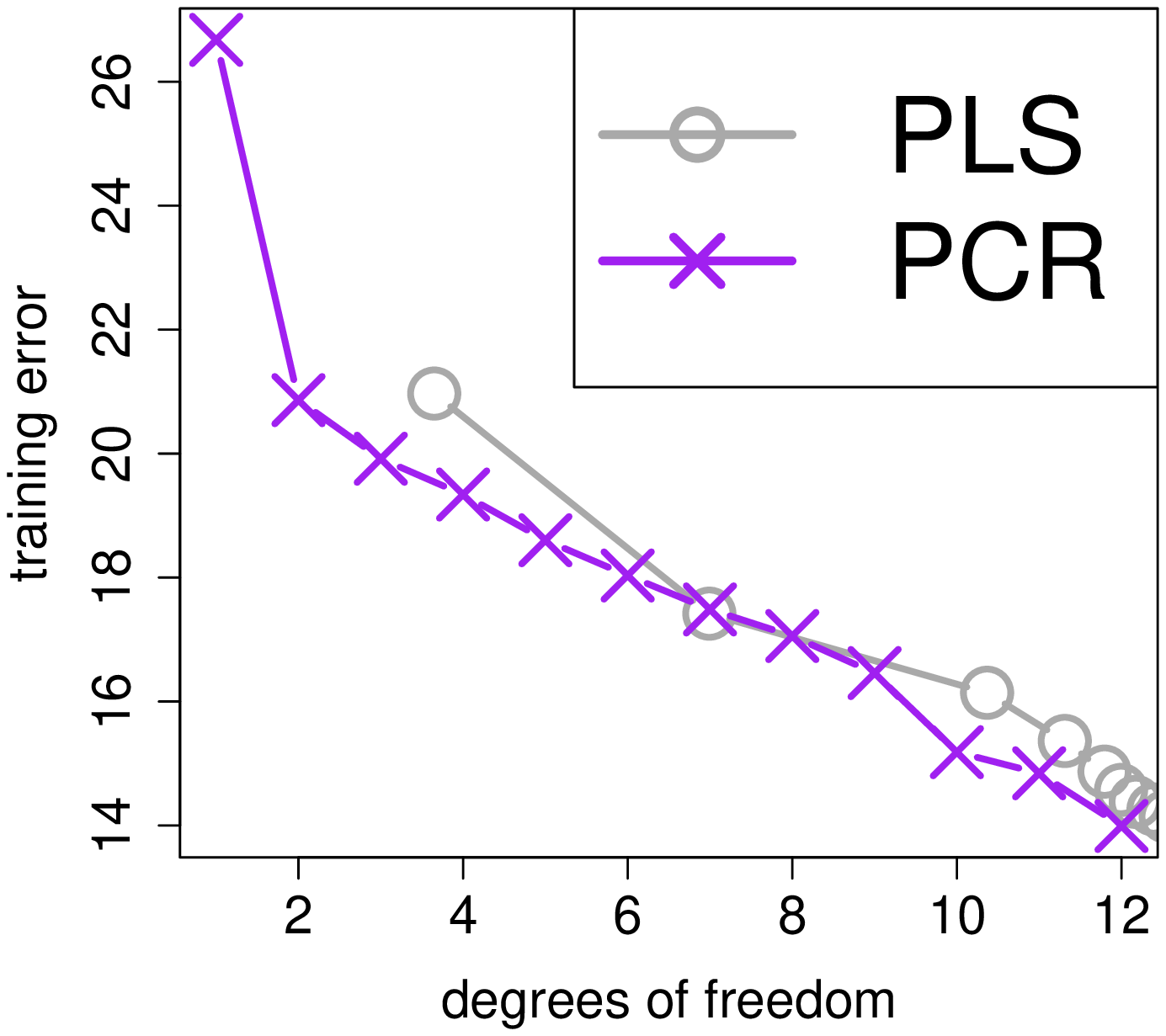}}
  \caption{Training error of PLSR and PCR.  Left: Training error as a function of the number of components. Right: Training error as a function of the Degrees of Freedom. }
  \label{fig:res}
\end{figure}

Further, it is known \citep{de93} that for a fixed number of components, PLSR obtains a lower approximation error
$\| \widehat \y_{ols} - \widehat \y_m \|^2 /n$ than PCR. Here, $\widehat \y_{ols}$ are the fitted values obtained by ordinary least squares. This implies that for a fixed number of components, the training error $\|  \y - \widehat \y_m \|^2 /n$ of PLSR is lower than the one of PCR. This is illustrated in the left plot of Figure \ref{fig:res}. Here, we plot the mean training error for PLSR and PCR over all 50 runs of the experiments as a function of the number of components.

The lower approximation error is sometimes used as an argument in favor of PLSR, as it often leads to the selection of fewer components compared to PCR (as illustrated in Figure \ref{fig:regression}). While this can be advantageous in problems where the latent components are used for model interpretation (e.g. for the visualization of the data in terms of a 2-D or 3-D plot), we stress again that the number of components do not capture the intrinsic model complexity, and that models with the same Degrees of Freedom should be compared instead. In the right plot of Figure \ref{fig:res}, we therefore plot the mean training error of PLSR and PCR as functions of the corresponding Degrees of Freedom. The plot shows that the difference of the approximation error is marginal in terms of Degrees of Freedom. Furthermore, the plot illustrates that PLSR concentrates a lot of components on regions with low training error and high complexity: All PLS models (with $m\geq 1$) consume at least $4$ Degrees of Freedom, and  for $m\geq 3$, the Degrees of Freedom already exceed $10$. Compared to PCR, this might be disadvantageous in situations where the true underlying model has a low complexity, as PLSR models do not explore the full range of Degrees of Freedom.

To summarize, the direct comparison of model parameters for different regression methods  is in general not possible, because they are either on a different scale (e.g. for PLSR and Ridge Regression) or they are just seemingly on the same scale (e.g. for PLSR and PCR) and a direct comparison would lead to misleading results. Degrees of Freedom offer a valuable solution to this problem.

\subsection{First Derivative of the Lanczos Decomposition}\label{subsec:lanczos} %
We now present the first implementation of the Degrees of Freedom estimate. We extend the approaches that are proposed in  \cite{Denham9701} and \cite{Serneels0401}. There, the iterative formulation of the NIPALS algorithm is used to construct the  derivative of $\widehat \bbeta_m$. Instead, our algorithm   is based on the matrix decomposition defined by \eqref{eq:L}. Note that the matrix $\L$ is upper bidiagonal, i.e. $ l_{ij}=0$ for $i >j $ or $i <j-1$. The relationship \eqref{eq:L} defines a Lanczos decomposition  \citep{Lanczos5001} of $\X$, i.e. a decomposition into orthonormal matrices $\T$ and $\W$ and an upper bidiagonal matrix $\L$. For fixed $\y$ and $m$, the decomposition is unique. The Lanczos decomposition can be interpreted as the analogue to the singular value decomposition that is defined by Principal Components Analysis. For more details on the equivalence of PLSR and Lanczos decompositions, we refer the readers to \cite{Elden0401}.

We propose an implementation of PLSR that iteratively constructs the derivative of the projection operator $\PP_{T}$ based on the Lanczos decomposition. This has the advantage that we also obtain the derivative of the regression coefficients $\widehat \bbeta_m$, which can then be used to construct approximative confidence intervals \citep{Denham9701,Phatak0201}.

We proceed in three steps. First, we derive a fast recursive PLSR algorithm based on the Lanczos decomposition. This algorithm avoids the explicit deflation of $\X$ as in \eqref{eq:deflation}, and only depends on projections onto one-dimensional subspaces. Second, we determine the first derivative of one-dimensional projection operators \citep{KraBra07}, which are
\begin{eqnarray*}
\frac{\partial \left(\v/\|\v\|_{\S}\right)}{\partial \y}&:=& \frac{\partial \left(\v/\sqrt{\v^\top \S \v}\right)}{\partial \y}= \frac{1}{\|\v\|_{\S}} \left(\I_n - \frac{\v \v^\top \S}{\v^\top \v} \right) \frac{\partial \v}{\partial \y}\,,
\end{eqnarray*}
and
\begin{eqnarray*}
\frac{\partial \left(\v \v^\top \z\right)}{\partial \y}&=& \left(\v \z^\top + \v^\top \z \I_n \right)\frac{\partial \v}{\partial \y} + \v \v^\top \frac{\partial \z}{\partial \y}\,.
\end{eqnarray*}
Finally, we differentiate the recursive formulas of the Lanczos representation. Algorithm \ref{algo:lanczos_p} displays the result. Its derivation can be found in the appendix.

\begin{algorithm}
\caption{Derivative of the regression coefficients and Degrees of Freedom}
\begin{algorithmic}[1]\label{algo:lanczos_p}
\STATE{Input: centered and scaled data $\X$, $\y$, number $m$ of components}
\STATE{$n=$ number or examples}
\STATE{$\S= \left(\X^\top \X\right)/(n-1)$, $\s= \left(\X^\top \y\right)/(n-1)$}
\STATE{Initialization: $ \widehat \bbeta_0={\bf 0}_p$, $\left({\partial \widehat \bbeta_0}/{\partial \y}\right)={\bf 0}_{p \times n}$}
\FOR{$i=1,\ldots,m$}
\STATE{$\w_i=\s- \S \widehat \bbeta_{i-1}$}
 \STATE{\quad \quad $\left({\partial \w_i}/{\partial \y}\right)= \X^\top/(n-1) - \S \left({\partial \widehat \bbeta_{i-1}}/{\partial \y}\right)$}
\STATE{$\v_i=\w_i - \sum_{j=1} ^{i-1} \v_j \v_j ^\top \S \w_i$}
 \STATE{\quad \quad  $\left({\partial \v_i}/{\partial \y}\right)= \left({\partial \w_i}/{\partial \y}\right) - \sum_{j=1} ^{i-1} \left({\partial \v_j \v_j^\top \S \w_i}/{\partial \y}\right)$}
\STATE{ $\v_i=\sqrt{n-1} \v_i/\|\v_i\|_{\S}$}
\STATE{\quad \quad $\left({\partial \v_i}/{\partial \y}\right)=\sqrt{n-1} \partial \left(\v_i/\|\v_i\|_{\S}\right)/\partial \y$}
\STATE{ $\widehat \bbeta_i= \widehat \bbeta_{i-1} + \v_i \v_i^\top \s$}
\STATE{\quad \quad  ${\partial \widehat \bbeta_{i}}/{\partial \y}={\partial \widehat \bbeta_{i-1}}/{\partial \y} + \left({\partial \v_i \v_i^\top \s}/{\partial \y}\right)$}
\ENDFOR
\STATE{$\dof(m)=1 + \text{trace}\left(\X \partial \widehat \bbeta_{m}/{\partial \y}\right)$.}
\end{algorithmic}
\end{algorithm}
As we compute the derivative of the regression coefficients as well, we can estimate the covariance of the PLSR coefficients by using a first order Taylor approximation
$\widehat \bbeta_m\approx \left(\partial \widehat \bbeta_m/\partial \y\right) \y$,
which leads to
\begin{eqnarray*}
\widehat{\text{cov}}\left(\widehat \bbeta_m\right)&=& \widehat \sigma^2  \frac{\partial \widehat \bbeta_m}{\partial \y} \left( \frac{\partial \widehat \bbeta_m}{\partial \y}\right)^\top \,.
\end{eqnarray*}
Furthermore, we can use the first order Taylor expansion to construct an approximate hat-matrix for PLSR via
\begin{eqnarray}
\label{eq:hat_pls}
 \H_{m,\,\text{approx}}&=& \frac{\partial \widehat \y_m}{\partial \y}\,.
\end{eqnarray}
This matrix can be plugged into formula \eqref{eq:sigmastar} for the estimation of the noise level.
\subsection{Trace of the Krylov Representation}\label{subsec:krylov} %
The computation of the derivative of $\widehat \y_m$ in Subsection \ref{subsec:lanczos} involves repeated matrix-matrix-multiplications. For high-dimensional data, this can become very time-consuming. As we do not need the derivative itself for the Degrees of Freedom but only its trace, we reduce the computational load by cleverly rearranging the computation of the derivative.

To this end, we use a closed form expression of the fitted values $\widehat \y_m$ that is based on Krylov subspaces. We use the fact \citep{Hoskuldsson8801} that $\text{span}\left\{\t_1,\ldots,\t_m\right \}=\text{span}\left\{\K\y,\ldots,\K^m\y\right \}=:\mathcal{K}_m$, with $\K= \X \X^\top$ the $n \times n$ kernel matrix. The space on the right-hand side is called the Krylov subspace defined by $\K$ and $\K \y$. We use the explicit representation $\widehat \y_m= \overline{\y} + \PP_{\mathcal{K}_m}$ to compute its derivative. In \cite{Phatak0201}, a corresponding formula for $\widehat \bbeta_m$ is used to determine its approximate distribution. We extend this result to the derivative of the fitted values. Additionally, after computing the derivative, we apply the basis transformation $\B = \left(\langle \t_i,\K^j \y\rangle  \right) \in \mathbb{R}^{m \times m}$ to improve numerical stability. This yields the following result.
\begin{propo}\label{pro:der}
Set
\begin{eqnarray*}
{\bf c}= \B^{-1} \T^\top \y \in \mathbb{R}^m&\text{ and }& \V=(\v_1,\ldots,\v_m )= \T \left(\B^{- 1}\right)^\top \in \mathbb{R}^{n \times m}\,.
\end{eqnarray*}
We have
\begin{eqnarray*}
 \frac{\partial \widehat \y_m}{\partial \y}&=&  \frac{1}{n} \I_n + \sum_{j=1} ^m c_j \left(\I_n -
\T \T ^\top \right) \K^j + \sum_{j=1} ^m \v_j \left(\y - \widehat \y_m\right)^\top \K^j + \T \T^\top\,.
\end{eqnarray*}
\end{propo}
In contrast to the Lanczos representation (Subsection \ref{subsec:lanczos}), the representation in Proposition \ref{pro:der} is more convenient for the computation of the Degrees of Freedom, as its trace can be computed directly.
\begin{propo}\label{pro:dof}
The unbiased estimate for the Degrees of Freedom of PLSR with $m$ components equals
\begin{eqnarray*}
\widehat {\operatorname{DoF}}(m)&=& 1+ \sum_{j=1} ^m c_j \operatorname{trace}\left(\K^j\right)  - \sum_{j,l=1} ^m  \t_l ´^ \top \K^j \t_l + \left(\y - \widehat \y_m\right)^\top \sum_{j=1} ^m   \K^j \v_j + m\,.
\end{eqnarray*}
\end{propo}
Hence, for the computation of the Degrees of Freedom of PLSR, we need a single fit of the PLSR algorithm that returns the matrix $\T$ of latent components. One can either use the original formulation of the NIPALS algorithm (Subsection \ref{subsec:pls}) or the Lanczos decomposition (Algorithm \ref{algo:lanczos_p}) without the computation of the derivative.
\section{Experiments}\label{sec:experiments} %
In this section, we evaluate the performance of the Degrees of Freedom estimate with respect to model selection for PLSR. We investigate the performance both with respect to prediction accuracy and model complexity (Subsection \ref{subsec:accuracy}), and  with respect to the estimation of the noise level $\sigma$ of the regression model (Subsection \ref{subsec:sigmahat}). Further, in Subsection \ref{subsec:stability}, we  discuss numerical issues and the computational efficiency of our proposed algorithms.

In our experiments, we consider the Bayesian Information Criterion. We conducted experiments with the Akaike Information Criterion and the generalized Minimum Description Length as well, but we found that our main observations on the difference between our Degrees of Freedom estimate and the naive approach $\dof(m)=m+1$ do not depend on the particular criterion. Hence, for the sake of clarity, we only report the results for the Bayesian Information Criterion.

We would like to stress that the primary goal of this section is not to advertise BIC in combination with our Degrees of Freedom estimate as a novel model selection criterion that is universally preferable to existing methods. Instead, we want to provide further empirical evidence that our estimate captures the intrinsic model complexity of PLSR reliably.

We use the  \texttt{ozone} data set as a starting point for our simulation study. As a preprocessing step, we scale all variables of the data set to lie in the interval [-1,+1]. The true regression function $f$  in equation \eqref{eq:reg1} is of the form
\begin{eqnarray*}
f(\x)&=& \sum_{j=1} ^d \beta_j  \phi_j(\x)\, \text{ with fixed basis functions } \,\phi_j(\x)= \text{exp}\left(- \|\x - \c_j\|^2\right)\,.
\end{eqnarray*}
The coefficients of the center points $\c_j \in \mathbb{R}^p$ are chosen uniformly from $[-1,+1]$. The regression coefficients $\beta_j$ are chosen randomly from a uniform distribution over $[1,3]$. We stress that the center points and hence the basis functions are fixed a-priori. This is to ensure that we are still in a parametric regression scenario, for which the Bayesian Information Criterion is suited. In the simulation study, we vary the number $d$ of basis functions from $10$ to $210$ in steps of $40$. We choose the variance $\sigma^2$ of the noise variable  such that the signal-to-noise-ratio equals $9$.  After the transformation of the initial data matrix $\X$  via the  basis functions $(\phi_1,\ldots,\phi_d)$, we obtain a $d$-dimensional data set, and are in the setting of a linear regression model.

We split the data into $50$ training and $153$ test points. The small training sample size allows us to consider high-dimensional settings, and the large test sample size ensures a reliable estimation of the test error. On the training data, we apply four different model selection criteria. (a) {\bf CV}: 10-fold cross-validation. (b) {\bf LANCZOS}: Bayesian Information Criterion with the Degrees of Freedom computed from the Lanczos decomposition (Algorithm \ref{algo:lanczos_p}). For the estimation of the noise level, we use equation \eqref{eq:sigmastar} with the approximate hat-matrix defined in \eqref{eq:hat_pls}. (c) {\bf KRYLOV}: Bayesian Information Criterion with the Degrees of Freedom computed from the Krylov representation (Proposition \ref{pro:dof}). For the estimation of the noise level, we use equation \eqref{eq:sigma}. (d) {\bf NAIVE}: Bayesian Information Criterion with the naive Degrees of Freedom $\dof(m)=m+1$.  For the estimation of the noise level, we use equation \eqref{eq:sigma}. Note that LANCZOS and KRYLOV use the same Degrees of Freedom estimate, and only differ in the estimation of the noise level $\sigma\,$. As the computation of the Degrees of Freedom depends on two different implementations, their runtime differs. Further, for all four methods, we set the range of the number of components from $0$ to $30$.

For each of the four criteria, we measure the performance on the  hold-out set of size $153$ by computing the normalized mean squared error: We divide the mean squared test error by the mean squared test error of the trivial model, i.e.  the constant model equal to the mean of the training data. This normalization facilitates the comparison between different values of $d$.  The procedure is repeated $50$ times.

\subsection{Prediction Accuracy and Model Complexity}\label{subsec:accuracy}
\begin{figure}[htb]
 \centering{\includegraphics[width=0.48\textwidth]{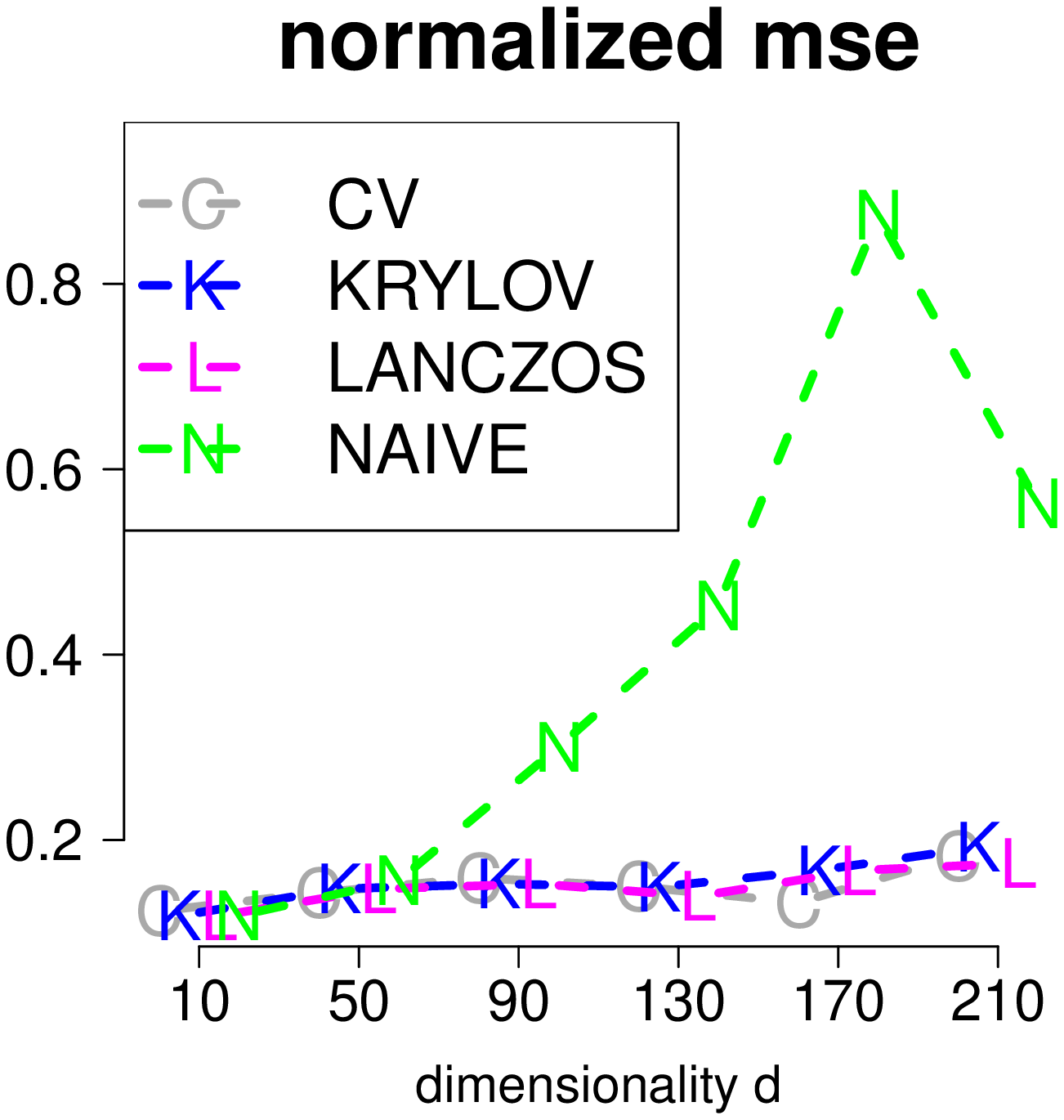} \includegraphics[width=0.48\textwidth]{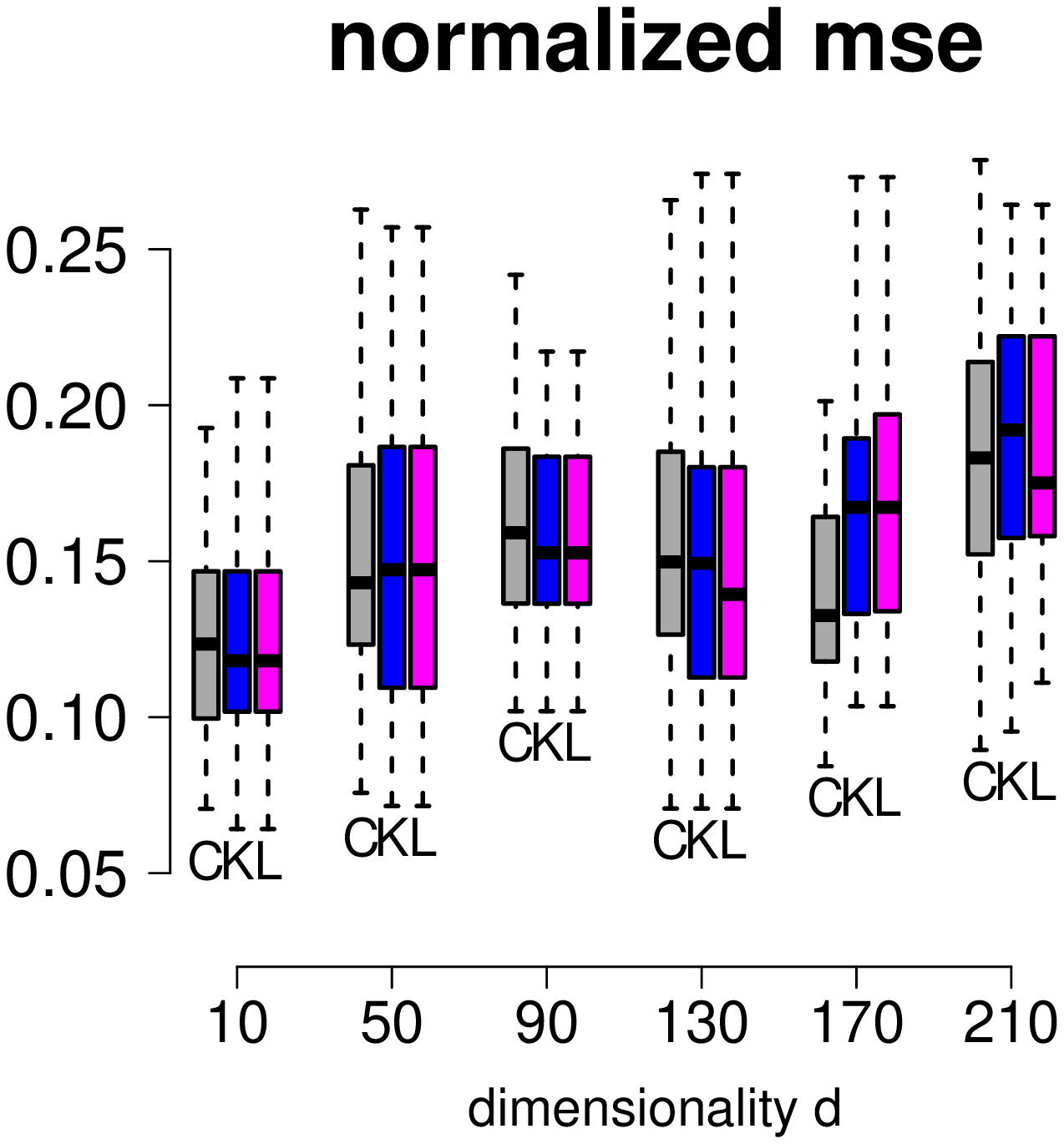}}
  \caption{Simulation results. Left: Median of the normalized test error for the four different model selection criteria. Right: Boxplot of the normalized test error for all methods except for the naive approach.}
  \label{fig:mse}
\end{figure}
We display the median and the boxplots of normalized test errors   in Figure \ref{fig:mse}. On average, our Degrees of Freedom estimate in combination with BIC shows a comparable prediction accuracy to cross-validation (right plot of Figure \ref{fig:mse}). The two different approaches for the computation of the Degrees of Freedom (KRYLOV and LANCZOS), which only differ in the estimation of $\sigma$, do not show any clear difference over the different simulation settings. Note however that LANCZOS is in general computationally more expensive. We refer the readers to Subsection \ref{subsec:stability} for more details on computation time.

\begin{figure}[htb]
 \centering{
 \includegraphics[width=0.48\textwidth]{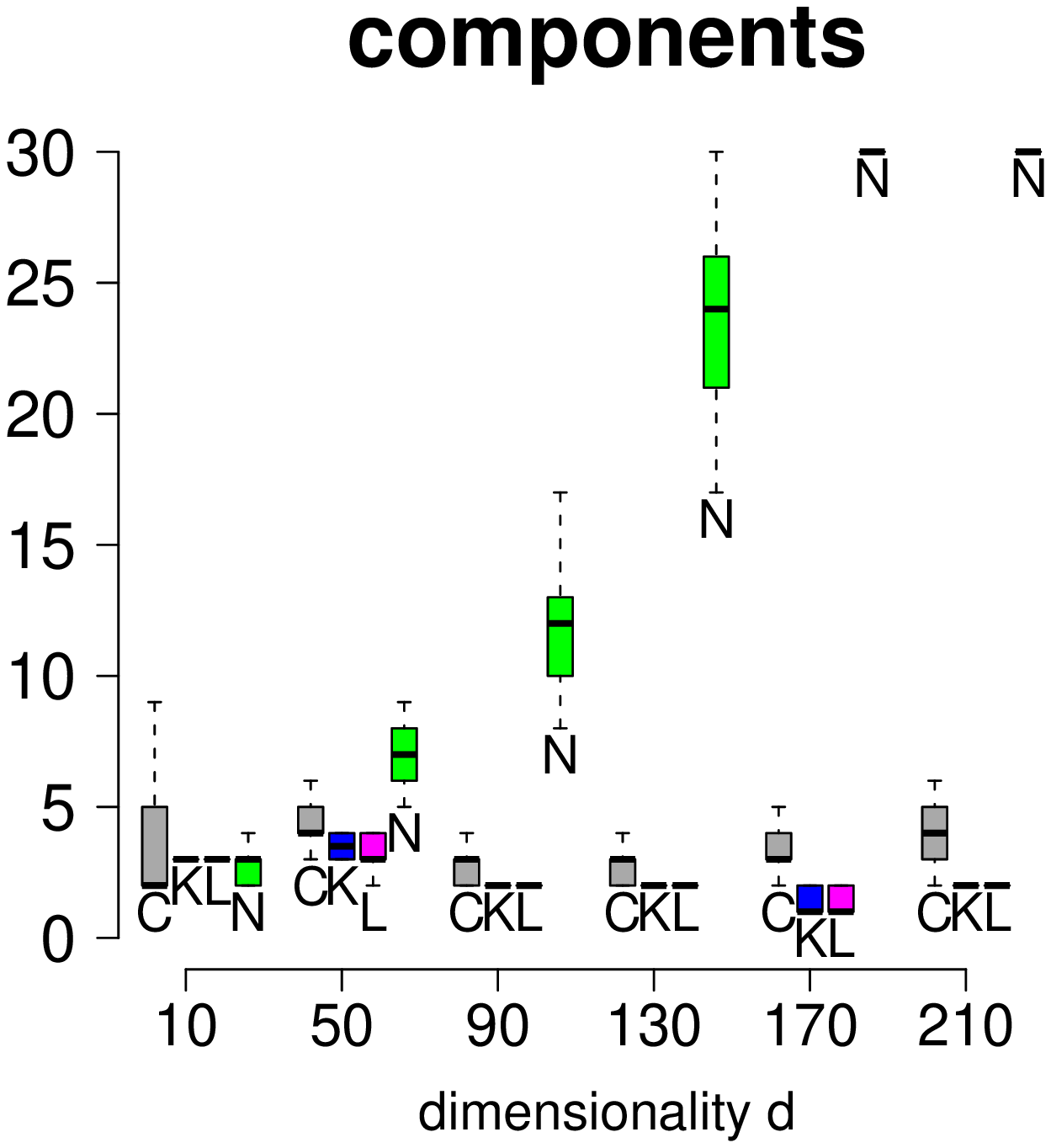}
 \includegraphics[width=0.48\textwidth]{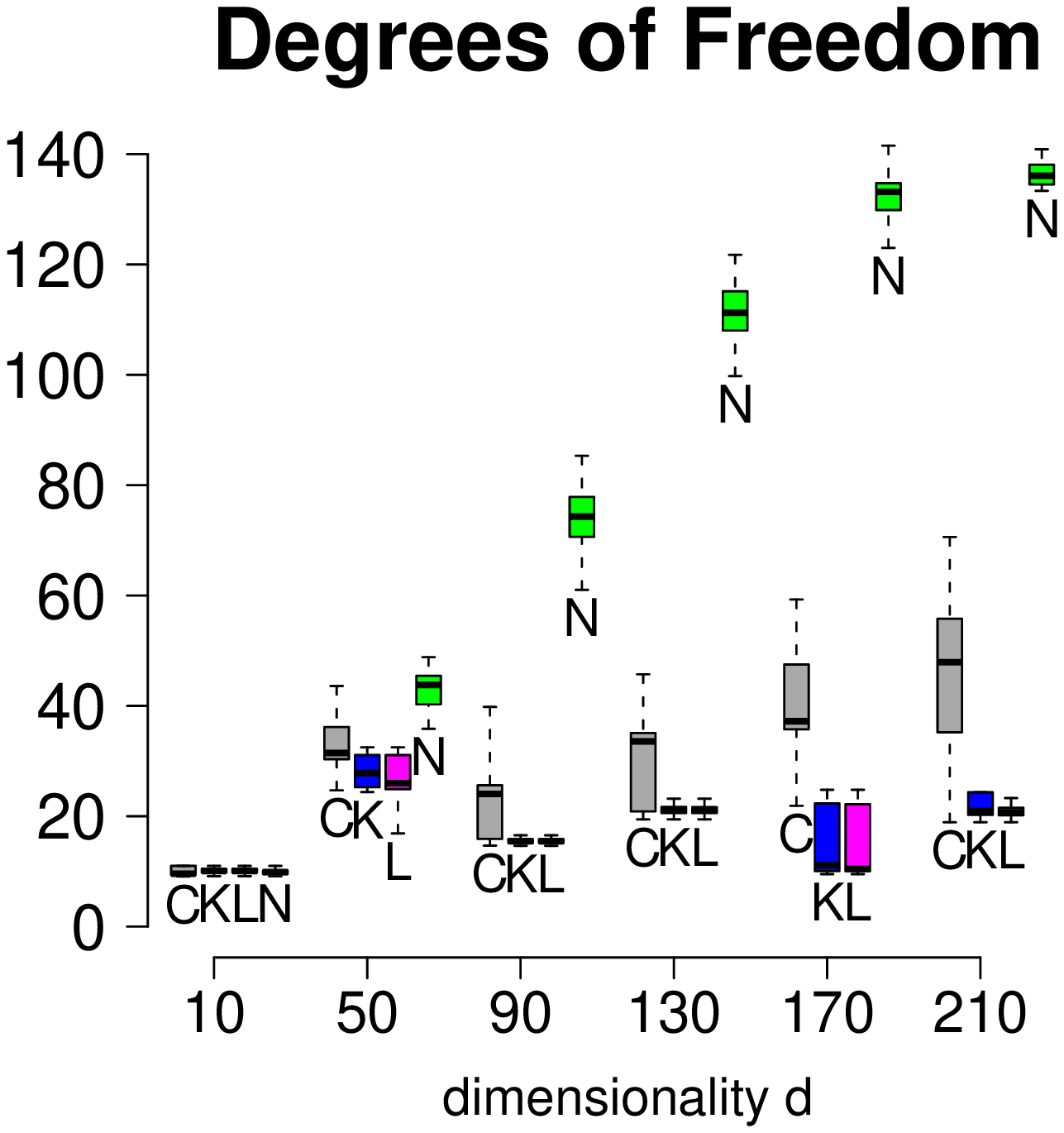}
 }
  \caption{Simulation results. Left: Number of selected components. Right: Degrees of Freedom of the selected model.}
  \label{fig:model}
\end{figure}

For the higher-dimensional scenarios ($d\geq 90$) in our simulation study, the naive approach leads to considerably worse results than the three other methods (left plot in Figure \ref{fig:mse}). Figure \ref{fig:model} illustrates that the naive approach tends to underestimate the true underlying complexity of the model. Compared to cross-validation and our $\dof$ estimate, it selects more components and Degrees of Freedom respectively.
A closer look at the simulation reveals that the selection of more components does not automatically lead to a higher test error. More precisely, for $d=50$ dimensions, the naive approach selects considerably more components compared to the three other methods, yet the prediction accuracy is on the same level. In contrast, for higher number of dimensions, the increased model complexity also affects the prediction accuracy. A possible explanation for this phenomenon is the following. For many moderate sized data sets, the test error first decreases sharply with the number of components, and then reaches a flat plateau for higher number of components. In this case, a more complex model can lead to a comparable prediction accuracy even with higher number of components.
\begin{figure}[htb]
 \centering{
 \includegraphics[width=0.48\textwidth]{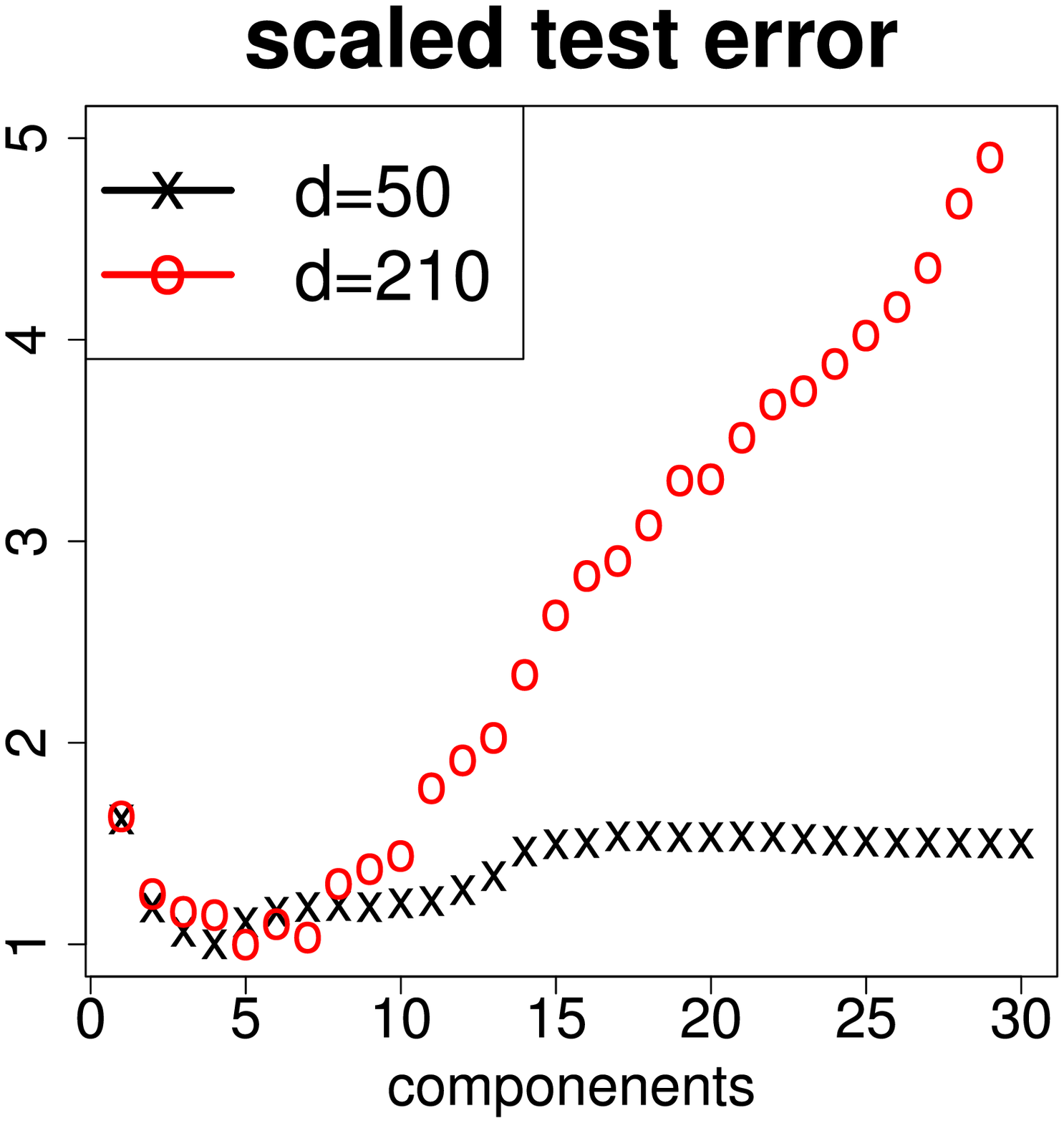}
 }
  \caption{Illustration: Scaled mean test error as a function of the number of components. For each value of $d$ (dimensionality), the test error is scaled such that its minimum value is $1$. }
  \label{fig:explanation}
\end{figure}
To underpin this point, we plot a scaled  test error for $d=50$ and $d=210$ as a function of the number of components (see Figure \ref{fig:explanation}). Here, we scale the test error such that its minimal value for a fixed $d$ is equal to 1. For $d=50$, the relative decrease in prediction accuracy is only moderate if we choose too many components, and the test error curve is flat. In contrast, for $d=210$, the relative increase of the test error is steep, and a selection of more components than the test-error optimal ones can lead to poor results.

\subsection{Estimation of the Noise Level $\sigma$}\label{subsec:sigmahat}
We now investigate the quality of the three different PLS based estimates for the noise level $\sigma$ that are obtained by KRYLOV, LANCZOS and NAIVE.
\begin{figure}[htb]
 \centering{\includegraphics[width=0.4\textheight]{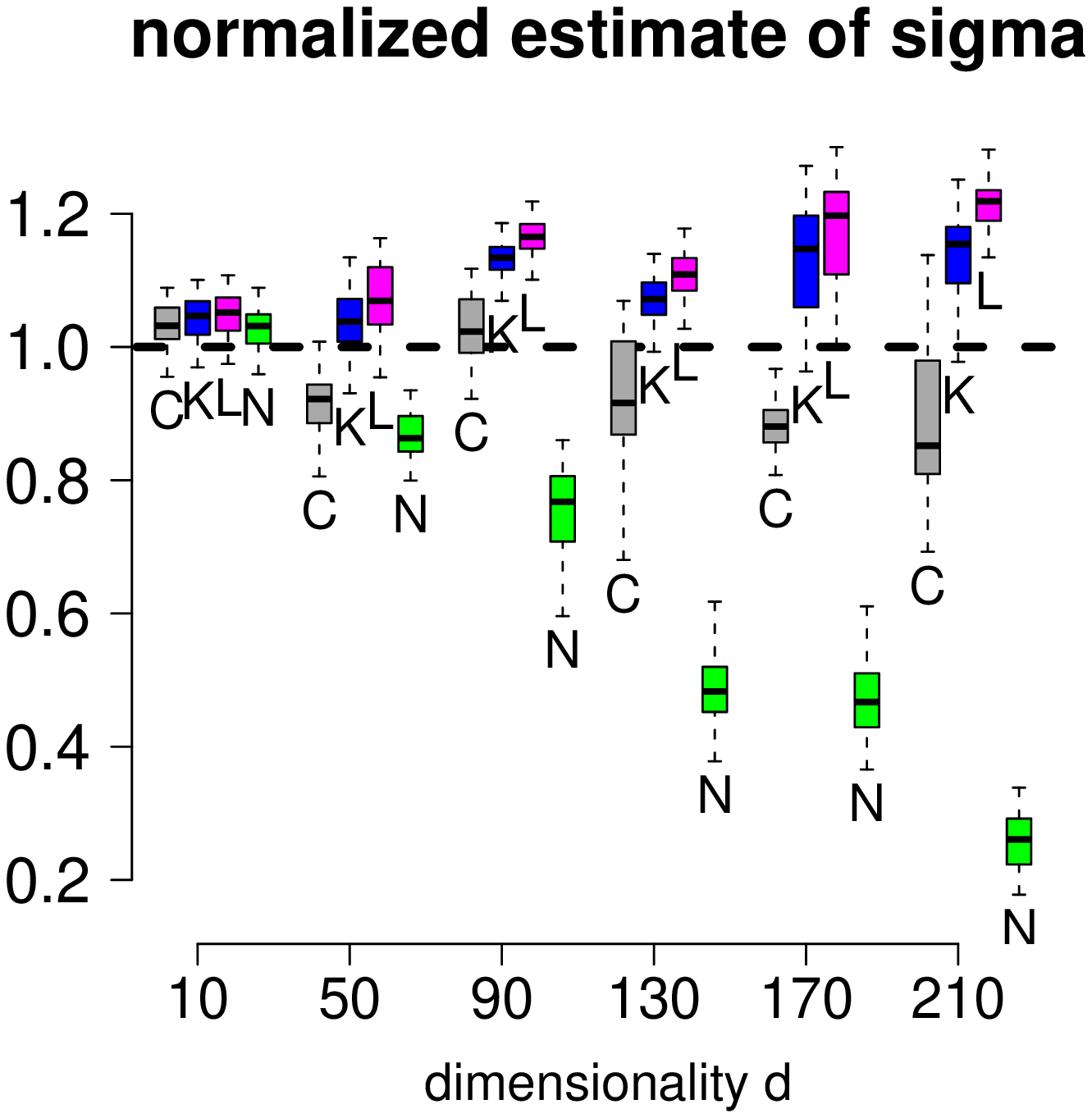}}
  \caption{Simulation results. Normalized estimated noise levels $\widehat \sigma/\sigma$. The dashed horizontal line indicates the theoretical optimum. }
  \label{fig:sigmahat}
\end{figure}

Figure \ref{fig:sigmahat} displays normalized estimates $\widehat \sigma$ obtained on the $50$ training samples: We divide each estimate by the true noise level $\sigma$ that is determined by the signal-to-noise ratio (which is set to $9$). The dashed horizontal line indicates the theoretical optimum of $1$. We observe that KRYLOV and LANCZOS slightly overestimate the noise level, which would in turn lead to slightly too conservative confidence intervals when used in inference problems. In contrast, the naive approach underestimates $\sigma$, and due to the complex models that it selects for higher dimensions (see Subsection \ref{subsec:accuracy}), the quality of the estimate deteriorates.
\subsection{Numerical Stability and Runtime}\label{subsec:stability}
As explained in Subsection \ref{subsec:lanczos}, the sparse structure of the matrix $\L$ defined in \eqref{eq:L} allows us to derive a fast iterative algorithm for PLSR and its derivative (Algorithm \ref{algo:lanczos_p}). In practice, we observe that the sparsity leads to numerical problems: After a certain number of components, the latent
components $\t_i$ are not mutually orthogonal anymore. This typically affects the computation of the Degrees of Freedom as well and leads to implausible results (e.g. negative Degrees of Freedom). In \cite{KraBra07}, the sparse structure of $\L$ is used and an additional stopping criterion is imposed to ensure that the latent components are orthogonal. However, this algorithm can stop very early. Therefore, we use equation
\eqref{eq:vi2} in the appendix that requires little additional computation time but ensures stability.

In some of the data, we observe that for a rather large number of components, both implementations for the Degrees of Freedom return negative Degrees of Freedom.  This indicates a  numerical problem. Therefore, in our experiments, we set the maximum number of components to $m_{*}$ if we observe negative Degrees of Freedom for $m_{*}+1$ components.

\begin{figure}[htb]
 \centering{\includegraphics[width=0.48\textwidth]{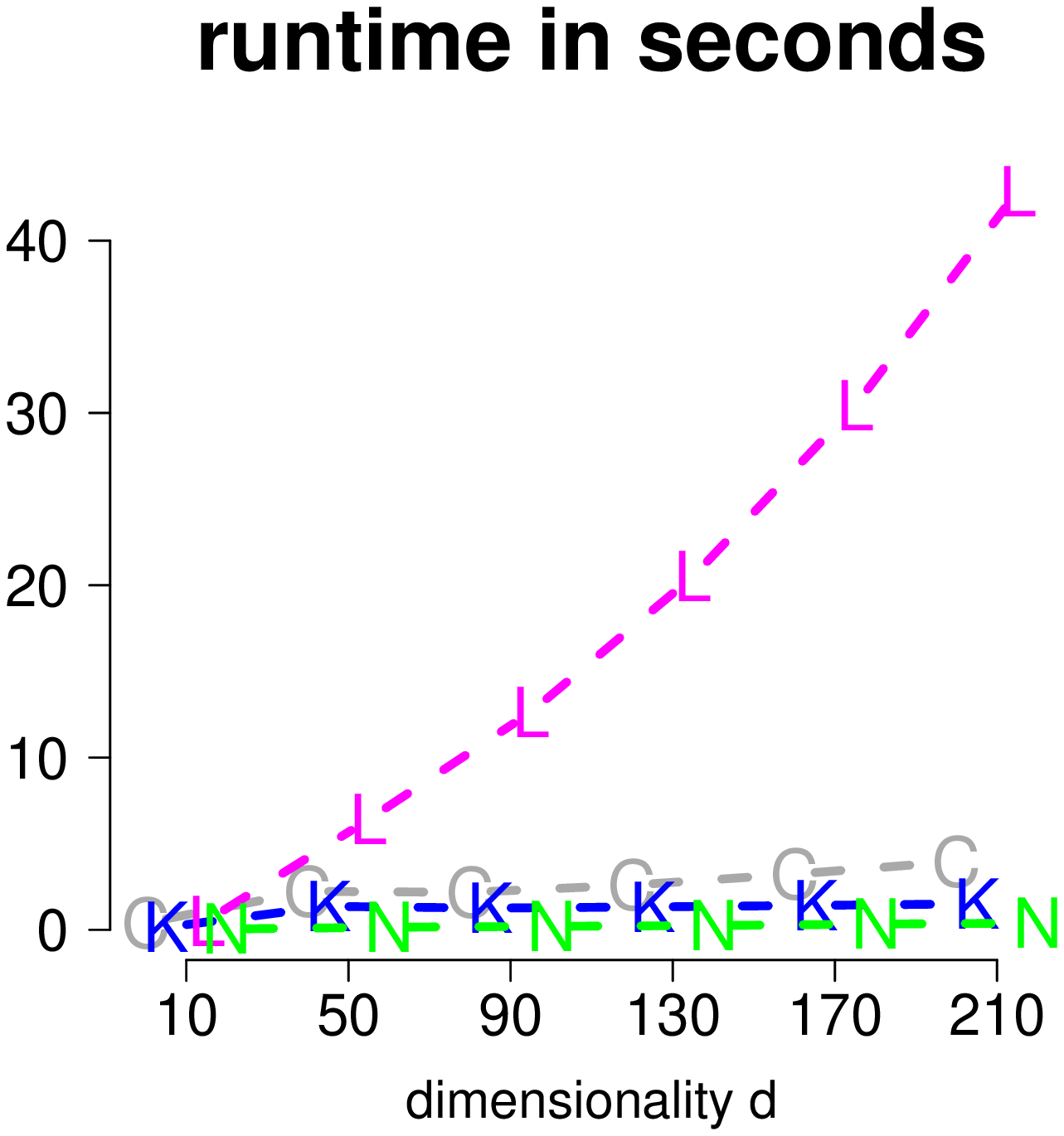} \includegraphics[width=0.48\textwidth]{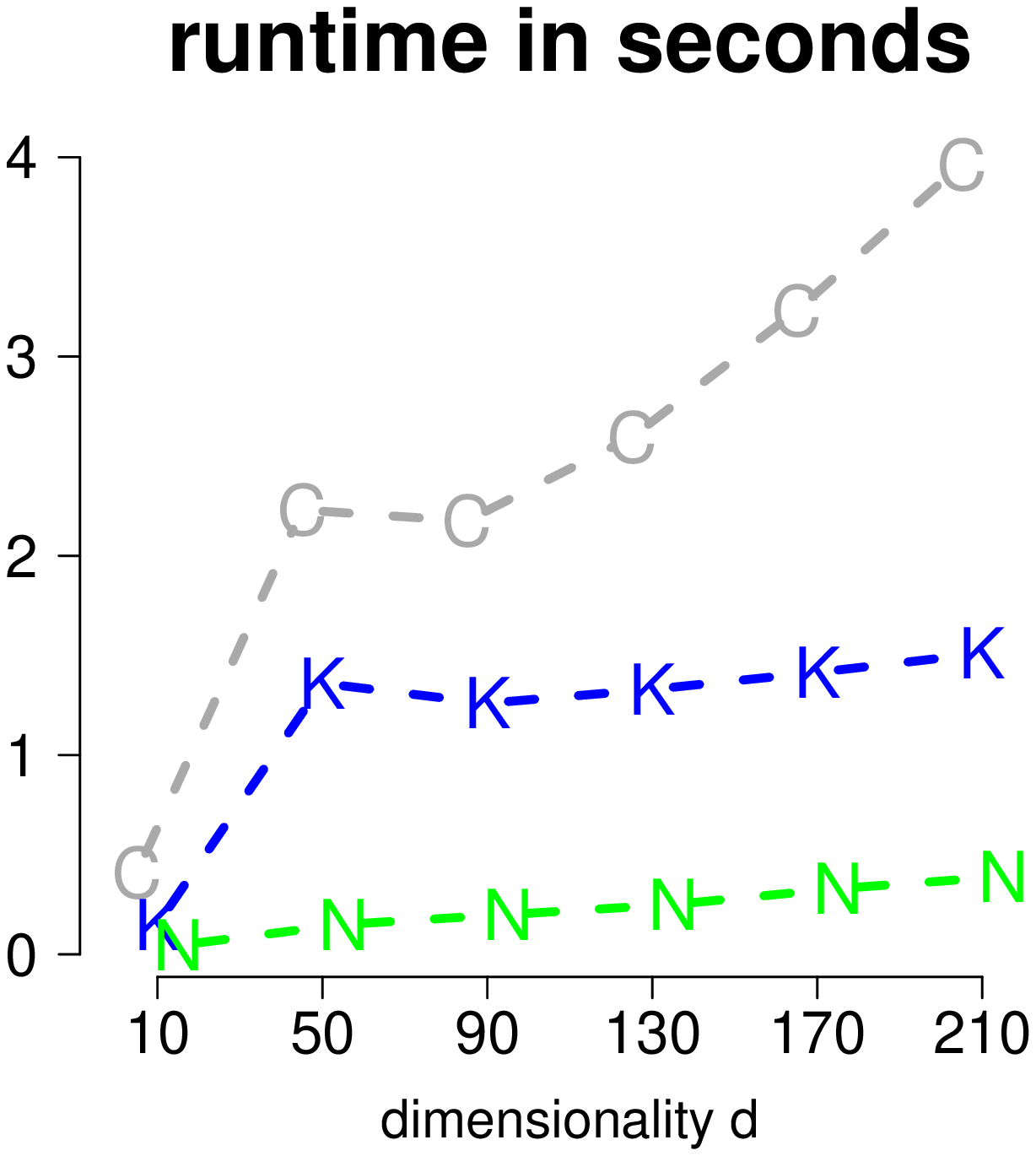}
 }
  \caption{Simulation results. Left: Median runtime in seconds for the four different methods. Left: Median runtime in seconds for all methods except for LANCZOS. }
  \label{fig:time}
\end{figure}

Finally, in Figure \ref{fig:time}, we compare the runtime of the four different methods in our simulation study. While the absolute values are low, the four criteria show clear differences. The Lanczos decomposition is by far the slowest approach, as it first computes the derivative of the PLSR fit before computing the Degrees of Freedom. With respect to runtime, the algorithm based on the Krylov representation is therefore preferable, if the explicit derivative is not needed. It is also faster than 10-fold cross-validation. The naive approach is always the fastest method, as it only requires a single run of the PLSR algorithm and no additional computation of the Degrees of Freedom.

\section{Discussion and Conclusion} %
Our findings show that typically, the Degrees of Freedom are higher for data sets with predictor variables that have low correlation and that each PLSR component consumes \emph{more} than one Degree of Freedom. This confirms the longstanding conjecture that $\dof(m)\geq m+1$. This result may not come as a big surprise: For a fixed number of components, PLSR is less biased than Principal Components Regression \citep{de93}. This decrease in bias is balanced by an increased complexity in terms of Degrees of Freedom.

On average, the Degrees of Freedom of PLSR in combination with information criteria yield  a similar prediction accuracy when compared to cross-validation. The naive approach that defines the Degrees of Freedom as the number of components plus one selects more complex models, which can in turn lead to worse prediction accuracy. This also effects the estimation of the noise level $\sigma$ of the regression model. While our approach slightly overestimates $\sigma$, the naive approach yields estimates that are considerably biased downwards. From the computational point of view, the implementation based on the Krylov representation is preferable to the Lanczos-based algorithm, and -- depending on the number $k$ of splits -- is faster than $k$-fold cross-validation.

In this paper, we applied the Degrees of Freedom estimate to the selection of the optimal number of PLSR components. It is possible to extend our framework to penalized PLSR \citep{Goutis96,Reiss07,KraBouTut2008a}, where an additional smoothing parameter has to be selected. The derivation of the Degrees of Freedom can be adapted accordingly.

The two implementations for the Degrees of Freedom capitalize on the close connection between PLSR and methods from numerical linear algebra, namely the Lanczos decompositions and Krylov subspace approximations. Apart from the computational advances that are pointed out in this paper, this connection is very fruitful to analyze statistical properties of PLSR in a concise way. Recent results on the correspondence of penalized PLSR  to preconditioning \citep{KraBouTut2008a} and on  the prediction consistency of PLSR \citep{BlaKra10,nips}  underpin the potential of this connection. We strongly believe that the interplay
between numerical linear algebra and PLSR
will further stimulate the field of statistics.

\paragraph{Acknowledgement} This work is funded in part by the FP7-ICT Programme of the European Community, under the PASCAL2 Network of Excellence, ICT-216886, and by the Funding Program for World-Leading Innovative R\&D on Science and Technology (FIRST).

\vskip 0.2in
\bibliographystyle{apalike}

\appendix
\section{Derivation of Algorithm \ref{algo:lanczos_p}}
The weight vector $\w_i$ can be rewritten as $\w_i= \X^\top \left(\y - \widehat \y_{i-1}\right)\propto  \s- \S \widehat \bbeta_{i-1}\,.$ We define the ``pseudo''-weight vector $\v_i$ via $\t_i= \X_i \w_i=:\X \v_i\,.$ Using the fact that the matrix $\L=\T ^\top \X \W$ defined in \eqref{eq:L} is upper-triangular, we yield
\begin{eqnarray}
\label{eq:vi}
\v_i&=&\w_i -  \v_{i-1} \v_{i-1} ^\top \S \w_i \\
\label{eq:vi2} &&\quad - \sum_{j=1} ^{i-2} \v_j \v_j ^\top \S \w_i\,.
\end{eqnarray}
Note that $\L$ is in fact also upper-diagonal, hence the term in second line is $0$. However, to ensure numerical stability, we include the term in our computation. The normalization of $\t_i$ to unit length corresponds to
\begin{eqnarray*}
\v_i&\leftarrow& \frac{\sqrt{n-1}}{\|\v_i\|_{\S}} \v_i=\frac{\sqrt{n-1}}{\sqrt{\v_i ^\top \S \v_i }} \v_i\,.
\end{eqnarray*}
It follows that $\widehat \bbeta_i= \widehat \bbeta_{i-1} + \v_i \v_i ^\top \s\,.$
\section{Description of the Data Sets}
\paragraph{\texttt{kin (fh)}} Simulation of the forward dynamics of an eight link all-revolute robot arm. The $32$ predictor variables correspond to positions of joints and to twist angles, length and offset distance for links. The task is to predict the distance of the end-effector from a target.  The problem is fairly linear (f) and contains a high amount of noise (h). The data is available at the delve-repository \url{http://www.cs.toronto.edu/~delve/}.

\paragraph{\texttt{ozone}} Los Angeles ozone pollution data 1976. The $12$ predictor variables contain the date of the measurement and information on wind speed, humidity, temperature etc. The task is to predict the daily maximum one-hour-average ozone reading. The original data contains missing values. From the $366$ examples, we use the $203$ examples with no missing values. The data is provided by the R-package `mlbench' \citep{mlbench}.

\paragraph{\texttt{cookie}} Quantitative NIR spectroscopy for dough piece. A Near Infrared reflectance spectrum is available for each dough piece. The spectral data consist of 700 points measured from 1100 to 2498 nanometers (nm) in steps of 2 nm. The task is to predict the percentage of fat. The data is first analyzed in \cite{Brown01,Osborne84}. The data set provided by the R-package `ppls' \citep{ppls} also contains the percentage of sucrose, dry flour, and water.

\end{document}